\documentclass[iicol,pdflatex,sn-mathphys-num]{sn-jnl}% Math and Physical Sciences Numbered Reference Style

\usepackage{graphicx}%
\usepackage{multirow}%
\usepackage{amsmath,amssymb,amsfonts}%
\usepackage{amsthm}%
\usepackage{mathrsfs}%
\usepackage[title]{appendix}%
\usepackage{xcolor}%
\usepackage{textcomp}%
\usepackage{manyfoot}%
\usepackage{booktabs}%
\usepackage{algorithm}%
\usepackage{algorithmicx}%
\usepackage{algpseudocode}%
\usepackage{listings}%
\usepackage{orcidlink}
\theoremstyle{thmstyleone}%
\theoremstyle{thmstyletwo}%

\theoremstyle{thmstylethree}%

\makeatletter
\newcommand{\sncollaboration}[1]{%
  \g@addto@macro\artauthors{%
    \par\vspace{0.3em}%
    {\normalfont\unboldmath\centering\large #1\par}%
    \vspace{0.3em}%
    \normalfont\unboldmath
  }%
}
\makeatother

\begin{document}

\noindent\makebox[\textwidth][r]{%
  \begin{minipage}{0.35\textwidth}
  \raggedleft
  \vspace{1cm}
 \hfill Belle II Preprint 2026-013\\
 \hfill KEK Preprint 2026-9
  \end{minipage}
}

\title[Article Title]{Measurement of mixing-induced \textit{CP} violation in~the~decay~$B^0 \to \pi^0 \pi^0$}

%author list
%%% Paper:    Measurement of mixing-induced CP violation in the $B^0 \to \pi^0 \pi^0$ decay
%%% Journal:  PhysRev
%%% Contacts: M. Dorigo, S. Raiz, D. Tonelli, K. Amos, R. \v{Z}leb\v{c}\'{i}k
%%% ====================================================================
%%% Use \input{pub110-orcid} to insert this material into your latex file.
  \author{M.~Abumusabh\,\orcidlink{0009-0004-1031-5425}} % 26883
  \author{I.~Adachi\,\orcidlink{0000-0003-2287-0173}} % 2590
  \author{K.~Adamczyk\,\orcidlink{0000-0001-6208-0876}} % 2239
  \author{A.~Aggarwal\,\orcidlink{0000-0002-5623-3896}} % 24463
% \author{L.~Aggarwal\,\orcidlink{0000-0002-0909-7537}} % 10083
% \author{P.~Ahlburg\,\orcidlink{0000-0002-9832-7604}} % 2367
% \author{H.~Ahmed\,\orcidlink{0000-0003-3976-7498}} % 11323
% \author{J.~K.~Ahn\,\orcidlink{0000-0002-5795-2243}} % 7423
  \author{Y.~Ahn\,\orcidlink{0000-0001-6820-0576}} % 14363
  \author{H.~Aihara\,\orcidlink{0000-0002-1907-5964}} % 2223
  \author{M.~Akdag\,\orcidlink{0009-0004-3728-1077}} % 27563
  \author{N.~Akopov\,\orcidlink{0000-0002-4425-2096}} % 9443
  \author{S.~Alghamdi\,\orcidlink{0000-0001-7609-112X}} % 27804
  \author{M.~Alhakami\,\orcidlink{0000-0002-2234-8628}} % 28103
% \author{A.~Aloisio\,\orcidlink{0000-0002-3883-6693}} % 2194
% \author{A.~Alsharari\,\orcidlink{0000-0002-6993-1597}} % 27803
% ...
% 13403
% \author{S.~Zou\,\orcidlink{0000-0003-3377-7222}} % 19363
\sncollaboration{(The Belle II Collaboration)}

\abstract{
Measurements of charge--parity (\textit{CP}) violation in quark transitions provide stringent tests of the current theory and sensitive probes of particles or interactions beyond it. We report the first measurement of mixing-induced \textit{CP} violation in $B^0\to\pi^0\pi^0$ decays. The measurement uses 190 million \mbox{$\Upsilon(4S)\to B^{0}\overline{B}{}^0$} events collected with the Belle~II experiment at the SuperKEKB asymmetric-energy electron-positron collider. A new approach that exploits the quantum entanglement of the $B^{0}\overline{B}{}^0$ pair enables a decay-time-dependent analysis without reconstructing the  $B^0 \to \pi^0\pi^0$ decay position. We measure the mixing-induced \textit{CP}-violating coefficient to be \mbox{$S_{\pi^{0}\pi^{0}} = 0.61^{+0.75}_{-0.79}\,(\mathrm{stat}) \pm 0.11\,(\mathrm{syst})$}, attaining a  precision that would require a data set twenty times as large with a conventional analysis. We also measure the direct  \textit{CP}-violating coefficient to be \mbox{$C_{\pi^{0}\pi^{0}} = 0.05 \pm 0.28\,(\mathrm{stat}) \pm 0.07\,(\mathrm{syst})$}. 
These results substantially improve the constraints from $B \to \pi\pi$ decays on the \textit{CP}-violating phase~$\phi_2$ of the quark-mixing matrix using far less data than had previously been assumed necessary.
}

%\keywords{}
%%\pacs[JEL Classification]{D8, H51}
%%\pacs[MSC Classification]{35A01, 65L10, 65L12, 65L20, 65L70}

\maketitle

Symmetries are central to the fundamental laws of nature. Charge--parity (\textit{CP}) symmetry, the invariance of physics laws under the combined action of charge conjugation and parity inversion, maps particles into antiparticles in a mirror-reflected spatial configuration. To date, \textit{CP} violation has been observed only in weak interactions that change quark flavour and, within the Standard Model of particle physics, is attributed to a single irreducible phase of the Cabibbo--Kobayashi--Maskawa (CKM) quark-mixing matrix~\cite{10.1143/PTP.49.652}. Its fundamental origin, however, remains unknown. Most importantly, studies of \textit{CP} violation  provide sensitive probes of physics beyond the Standard Model, by revealing or constraining additional \textit{CP}-violating phases induced by new particles or interactions.

An important element of these studies is the measurement of the \textit{CP}-violating phase
$\phi_2 \equiv \arg\!\left[-(V_{td}V_{tb}^{*})/(V_{ud}V_{ub}^{*})\right]$ (also denoted as~$\alpha$), where $V_{ij}$ is the CKM-matrix element connecting the up-type quark $i$ to the down-type quark~$j$. Experimentally, $\phi_2$ is determined from specific $B$-meson decays to light mesons---such as $B^{0}\to\pi^{+}\pi^{-}$, $B^{+}\to\pi^{+}\pi^{0}$, and \mbox{$B^{0}\to\pi^{0}\pi^{0}$}---by combining their decay-time-dependent and \mbox{-independent} amplitudes through isospin relations~\cite{Gronau:1990ka,Charles:2017evz}. Charge-conjugate modes are implied throughout unless explicitly stated otherwise. Complementary constraints are obtained from analogous isospin analyses of the $B\to\rho\rho$ and $B\to\rho\pi$ decays.  

The $B \to \pi\pi$ system is experimentally simpler because $\rho$ mesons are broad resonances with more complex kinematic properties and are harder to distinguish from background. In $B \to \pi\pi$ decays, however, the $\phi_{2}$ determination has long been hindered by multiple discrete ambiguities in the allowed $\phi_2$ values due to the absence of a key input, the coefficient $S_{\pi^0\pi^0}$ encoding mixing-induced \textit{CP} violation in the $B^{0}\to\pi^{0}\pi^{0}$ decay. As a consequence, the $B\to\rho\rho$ system has provided the leading constraints on $\phi_2$ to date.  

A measurement of $S_{\pi^0\pi^0}$ would substantially reduce the ambiguities in the $B \to \pi \pi$ system. This would restore $B \to \pi \pi$ as a competitive probe of $\phi_2$, enabling independent, comparably precise determinations across channels affected by different experimental and theoretical uncertainties, and ultimately accelerating progress on the determination of $\phi_2$.
However, a measurement of $S_{\pi^0\pi^0}$ with the conventional approaches  would require data samples an order of magnitude larger than currently available.

We report the first measurement of $S_{\pi^0\pi^0}$ using a new approach, which substantially improves the determination of $\phi_2$ from $B \to \pi\pi$ decays. 
We use data collected with the Belle~II experiment at the asymmetric-energy electron-positron SuperKEKB collider~\cite{AKAI2018188,abe2010belleiitechnicaldesign}. Belle~II is a large-solid-angle 1.5\,T solenoidal spectrometer comprising silicon vertex detectors, charged-particle tracking and identification systems, and an electromagnetic calorimeter, surrounded by a muon and $K_{L}^0$ detector~\cite{abe2010belleiitechnicaldesign,Miyabayashi_2020}. 

At Belle~II, neutral $B$ mesons are produced near kinematic threshold, in the process \mbox{$e^+e^- \to \Upsilon(4S)\to B^{0}\overline{B}{}^0$}, as a quantum-entangled $B^{0}\overline{B}{}^0$ pair, and may undergo $B^{0}$--$\overline{B}{}^0$ mixing before decaying. For decays to a final state $f$ that is accessible to both $B^{0}$ and $\overline{B}{}^0$ mesons (such as $\pi^{0}\pi^{0}$), interference between the direct decay and the decay following mixing generates a time-dependent \textit{CP}-violating asymmetry
\begin{equation}
A_{\it C\!P}(t) = \frac{N(\overline{B}{}^0(t) \to f) - N(B^0(t) \to f))}{N(\overline{B}{}^0(t) \to f) + N(B^0(t) \to f))}\,,
\end{equation}
where $N$ denotes the yield and $t$  the proper decay time.  Experimentally, the asymmetry is accessed by reconstructing the signal decay and tagging the flavour of the companion meson (tag $B$), which identifies the signal meson as a $B^{0}$ or $\overline{B}{}^0$ at the reference time set by the tag-$B$ decay. The asymmetry is parameterised by coefficients $S_{f}$ and $C_{f}$, describing mixing-induced and direct \textit{CP} violation, respectively, and is conventionally extracted from the distributions of the decay-time difference $\Delta t$ between the signal- and tag-side decays. This quantity is inferred from the separation of the two decay positions (vertices) reconstructed from charged-particle trajectories (tracks). The \mbox{$B^{0}\to\pi^{0}\pi^{0}$} signal, however, is typically reconstructed from four photons produced in the dominant $\pi^{0}\to\gamma\gamma$ decay, which does not provide a signal vertex. Such a vertex could be obtained only in rare topologies, such as \mbox{$\pi^{0}\to e^{+}e^{-}\gamma$}. This renders $\Delta t$-based measurements feasible only with samples of about 10 billion $\Upsilon(4S)$ decays or more~\cite{Belle-II:2018jsg}.

\begin{figure}[t]
\centering
\includegraphics[width=\columnwidth]{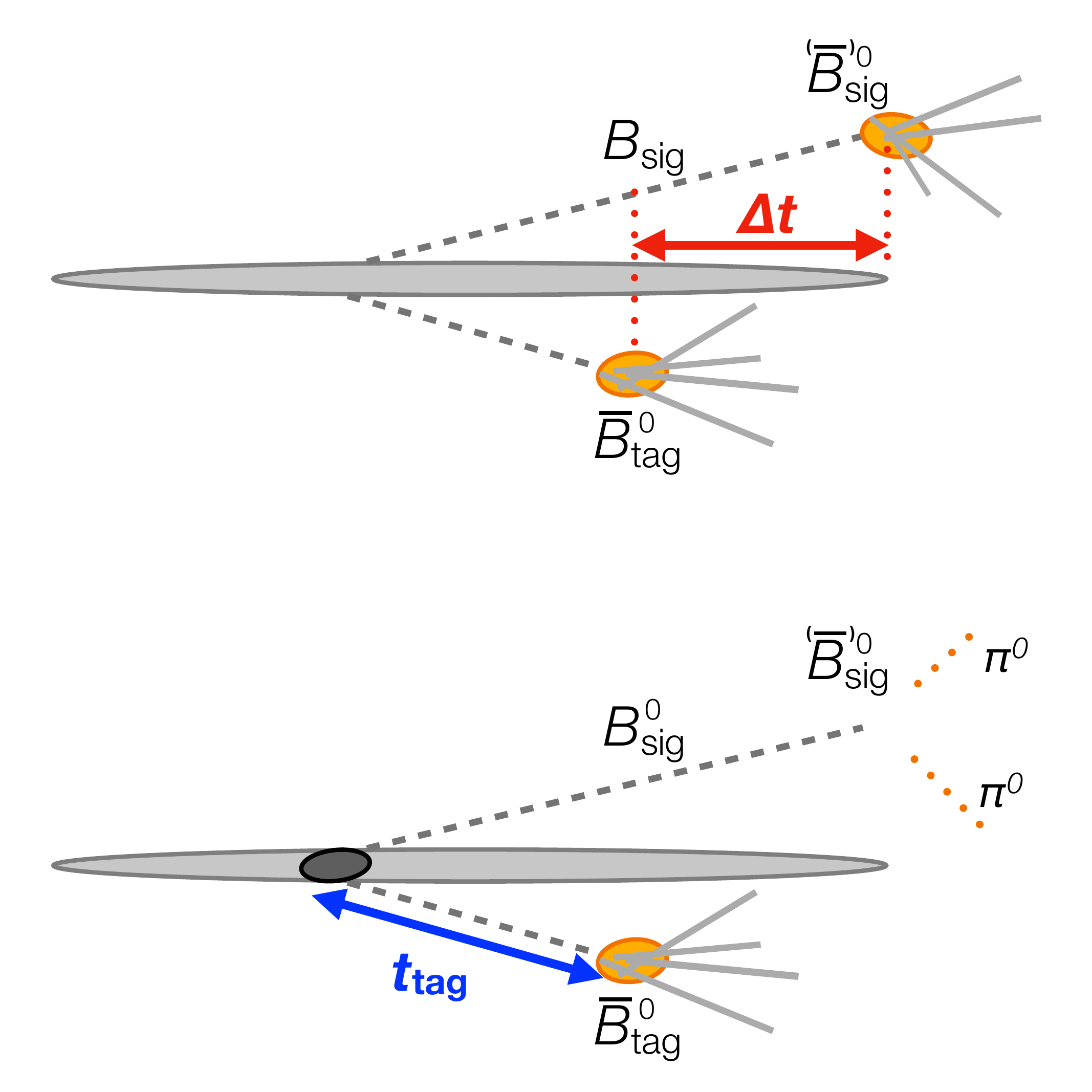}
\caption{Schematic illustration, not drawn to scale, of the conventional $\Delta t$ measurement and the $t_{\rm tag}$ method.  The $B$ mesons are produced in the luminous region (elongated grey ellipse) from a $\Upsilon(4S)$ decay and move from left to right along the boost direction. Top: in the conventional method, the decay vertices (orange ellipses) of both the signal $B^0$ meson ($B^0_{\rm sig}$) and the tag-$B$ meson ($\overline{B}{}{^0_{\rm tag}}$) are reconstructed from the observed tracks to measure the decay-time difference $\Delta t$ (red double arrow). Bottom: when $B^0_{\rm sig}$ decays into $\pi^0(\to \gamma\gamma)\pi^0(\to \gamma\gamma)$, its decay vertex  cannot be reconstructed. The common production point of the  $B^0\overline{B}{}{^0}$ pair (dark grey ellipse) in the luminous region and the decay vertex of the tag $B$ (orange ellipse) are reconstructed to measure the tag-$B$ decay time $t_{\rm tag}$ (blue double arrow). 
}
\label{fig:vertexDiagram}
\end{figure}

In this work, we use a much smaller sample of $190$ million $\Upsilon(4S)\to B^{0}\overline{B}{}^0$ decays collected between 2019 and 2022, corresponding to an integrated luminosity of 365\,fb$^{-1}$. The measurement is enabled by a novel method~\cite{PhysRevD.60.111301,rzr8-l6l8}, which is applied experimentally for the first time in this work.  Rather than using $\Delta t$, we study the time evolution of the system as a function of $t_{\rm tag}$, the time interval between the production of the tag $B$ meson and its decay.  A schematic comparison between the conventional $\Delta t$-based and  $t_{\rm tag}$-based approaches is shown in Fig.~\ref{fig:vertexDiagram}.  

Because the two $B$ mesons are quantum-entangled, the probability density $P(t_{\rm tag}, q)$ for a tag $B$ meson to decay at proper time $t_{\rm tag}$ with flavour $q$ retains information about decay-time-dependent \textit{CP} violation in the signal decay. The parameters $S_{\pi^{0}\pi^{0}}$ and $C_{\pi^{0}\pi^{0}}$ can therefore be determined without reconstructing the signal vertex, according to
\begin{equation}
\label{eq:prob}
\begin{split}
& P(t_{\rm tag}, q) \approx  \\
 & \quad  e^{-t_{\rm tag}/\tau} \bigg[ 1 - q\,D  \Big( S_{\pi^0\pi^0}\sin[\Delta m (t_{\rm tag}-\hat{t}\,)] \\
& \quad \quad \quad \quad \qquad +  C_{\pi^0\pi^0}\cos[\Delta m (t_{\rm tag}-\hat{t}\,)]
\Big)
\bigg] \,,
\end{split}
\end{equation}
where $\tau$ is the average $B^0$ lifetime; $\Delta m$ is the $B^{0}$--$\overline{B}{}^0$ flavour-mixing frequency; \mbox{$q=+1$ ($-1$)} is the $B^0$ ($\overline{B}{}^0$) flavour of the tag meson; $\hat{t} = \arctan (\Delta m \tau)/\Delta m\approx 1.294\,\rm{ps}$ is the decay time at which the mixing-induced term in the decay-rate asymmetry first vanishes; and \mbox{$D = 1/\sqrt{1 + (\tau\,\Delta m)^2}\approx 0.792$} is a damping of the asymmetry amplitude due to integrating over all possible signal decay times~\cite{PhysRevD.60.111301,rzr8-l6l8}. 

The method relies on an accurate determination of $t_{\rm tag}$, obtained from the displacement between the production point of the tag $B$ meson and its decay vertex~\cite{rzr8-l6l8}.
The decay vertex is obtained from reconstructed tracks, whereas the production point is inferred from the luminous region of the colliding beams and the kinematic constraints of the event.
A $t_\mathrm{tag}$ measurement is therefore feasible only if a sufficiently compact interaction region and excellent vertexing are available. Unlike first-generation $B$ factories, SuperKEKB has a micron-scale transverse luminous region, while the Belle~II pixel detector has a vertex resolution of a few tens of microns~\cite{Belle-IIDEPFET:2021pib}. Together, these enable a per-event estimate of $t_{\rm tag}$ with a resolution of about $1.5\,\mathrm{ps}$~\cite{rzr8-l6l8},  sufficient for time-dependent studies without signal vertexing. The nonzero beam crossing angle, through the transverse boost it induces, further improves the time resolution.

The analysis builds on the $B^{0}\to\pi^{0}\pi^{0}$ branching-fraction measurement obtained from the same data, which also determined $C_{\pi^0\pi^0}$ through a time-integrated asymmetry~\cite{BelleIIpi0pi0BR}.  We extend that analysis by incorporating $t_{\rm tag}$ and its per-event uncertainty $\sigma_{t_{\rm tag}}$, obtained following Ref.~\cite{rzr8-l6l8}, and extract $S_{\pi^0\pi^0}$ and $C_{\pi^0\pi^0}$ with a maximum-likelihood fit. Because the sensitivity to $S_{\pi^{0}\pi^{0}}$ depends directly on the modelling of the $t_{\rm tag}$ resolution, we determine the resolution parameters by measuring the $B^{0}$--$\overline{B}{}^0$ oscillation frequency in $B^0\to D^-\pi^+$ decays reconstructed in data.  We then validate the $t_{\rm tag}$ approach by reproducing the known \textit{CP}-violating coefficients $S_{J/\psi K^0_{\rm S}}$ and $C_{J/\psi K^0_{\rm S}}$  in the well-established $B^{0}\to J/\psi K^{0}_{\rm S}$ decay. 

The signal decay $B^{0}\to\pi^{0}\pi^{0}$ is reconstructed using only calorimeter information. Photon candidates are formed from energy deposits (clusters)  not associated with tracks and required to satisfy basic energy and timing criteria.
Data-driven corrections are applied to the photon energy. Pairs of photons with invariant mass consistent with \mbox{$\pi^{0}\to\gamma\gamma$} decays are $\pi^0$ candidates, and their momentum measurement is refined using a mass-constrained kinematic fit. A multivariate photon-quality selection based on cluster-energy, shape, and isolation observables suppresses background from beam-induced and misreconstructed photons~\cite{BelleIIpi0pi0BR}.  All tracks originating from the interaction region are assigned to the tag $B$ meson to reconstruct its decay vertex.

The background is dominated by high-energy $\pi^0$ mesons from $e^{+}e^{-}\to q\bar q$  production (continuum), where $q$ indicates a $u$, $d$, $s$, or $c$ quark. After reconstruction, this background exceeds the expected signal by roughly three orders of magnitude. The signal-to-background ratio is improved using event-shape observables that exploit the topological differences between continuum events, which are jet-like,  and the $B\overline{B}$ events, which are more isotropic. We train a boosted-decision-tree classifier on 27 such observables to discriminate signal from continuum.  Continuum is modelled using data collected at energies 60 MeV below the $\Upsilon(4S)$ mass peak, where $B\overline{B}$ production is kinematically forbidden; the sample size is about one-tenth of the on-resonance dataset. Signal is modelled using simulation. We optimise the requirement on the classifier output by minimising the average expected statistical variance on $S_{\pi^0\pi^0}$ as determined from repeating the measurement on ensembles of simulated bootstrap replicas~\cite{Efron1979}. The resulting selection rejects 98\% of the continuum while retaining 80\% of the signal.

Loose requirements, which retain high signal efficiency, are applied to observables that exploit the kinematic features of near-threshold production. These are the beam-energy-constrained mass $M_{\rm bc}$, which is the signal $B$ invariant mass with the $B$ energy replaced by one-half the centre-of-mass energy,  and the difference $\Delta E$ between the reconstructed $B$-candidate energy and one-half the centre-of-mass energy, both calculated in the $\Upsilon(4S)$ rest frame.

In addition to signal and continuum, the sample contains background from misreconstructed $B$ decays
($B \overline{B}$ background), whose yield is comparable to the expected signal. This background is dominated by $B^+\to\rho^+(\to\pi^+\pi^0)\pi^0$ decays, with smaller contributions from $B^0$ decays such as  $B^0\to K_S^0(\to\pi^0\pi^0)\pi^0$. Because these backgrounds are not fully reconstructed, they populate lower $\Delta E$ values than the signal.  Misreconstructed signal candidates, in which a $\pi^0$ from the tag $B$ decay is incorrectly associated with the signal, amount to less than 2\% of the expected signal yield. The fraction of events with multiple signal candidates is below 1\%; for these events, we select one candidate at random. 

For each event, the flavour of the tag $B$ meson is inferred with a multivariate flavour-tagging algorithm, which provides both a flavour decision $q$ and a probability $w$ for that decision to be wrong, as calibrated in control data~\cite{flavorTagger}.

We extract $S_{\pi^0\pi^0}$ and $C_{\pi^0\pi^0}$ with an extended maximum-likelihood fit that models the sample as a mixture of signal (including misreconstructed candidates), continuum, and $B\overline{B}$ background. The fit uses the unbinned distributions of seven observables $(\Delta E,\,M_{\rm bc},\,C_t,\,q,\,w_t,\,t_{\rm tag},\,\sigma_{t_{\rm tag}})$, where $C_t$ and $w_t$ are, respectively, the continuum-suppression classifier output and the wrong-flavour probability, both transformed to facilitate modelling~\cite{Fisher}. The observables $M_{\rm bc}$ and $C_t$ separate signal from continuum, while $\Delta E$ provides additional discrimination against $B\overline{B}$ background.

For each sample component, the probability density function (PDF) is factorised into a time-independent part describing the  $(\Delta E,\,M_{\rm bc},\,C_t)$ distributions and a time-dependent part describing the $(q,\,w_t,\,t_{\rm tag},\,\sigma_{t_{\rm tag}})$ distributions.  
The time-independent model follows the time-integrated analysis~\cite{BelleIIpi0pi0BR}, with small adjustments to account for the difference in the continuum-suppression selection.

The signal PDF for $t_{\rm tag}$ is constructed by including flavour-tagging imperfections in Eq.~\ref{eq:prob} and convolving the result with the $t_{\rm tag}$-resolution function. Only $S_{\pi^0\pi^0}$ and $C_{\pi^0\pi^0}$ are freely determined by the fit, while external physics inputs and flavour-tagging calibration parameters are Gaussian-constrained to either their known values or values measured in calibration samples~\cite{flavorTagger}.  The $B\overline{B}$ background is described with the same time-dependent model based on empirical nuisance parameters; the continuum time dependence is described directly from data with a flexible empirical model.

\begin{figure*}[t]
\centering
\includegraphics[width=160pt]{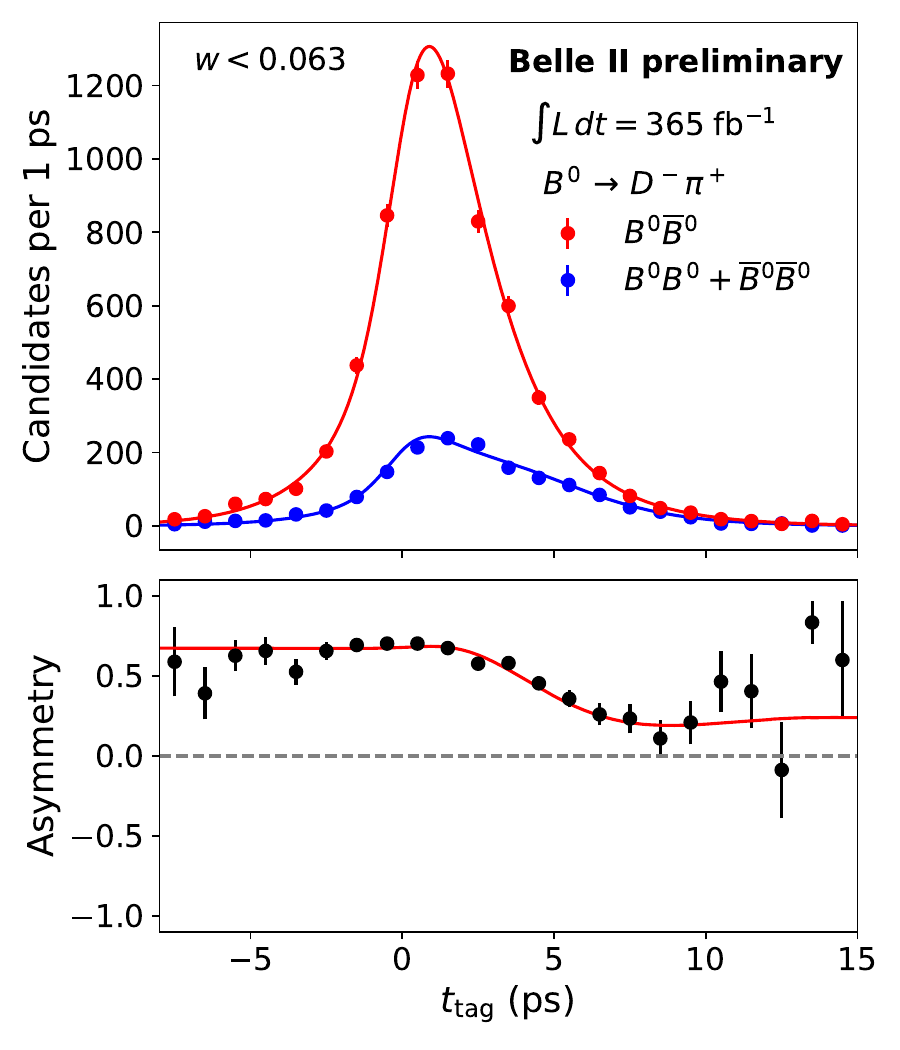}\hspace{2.5cm}
\includegraphics[width=160pt]{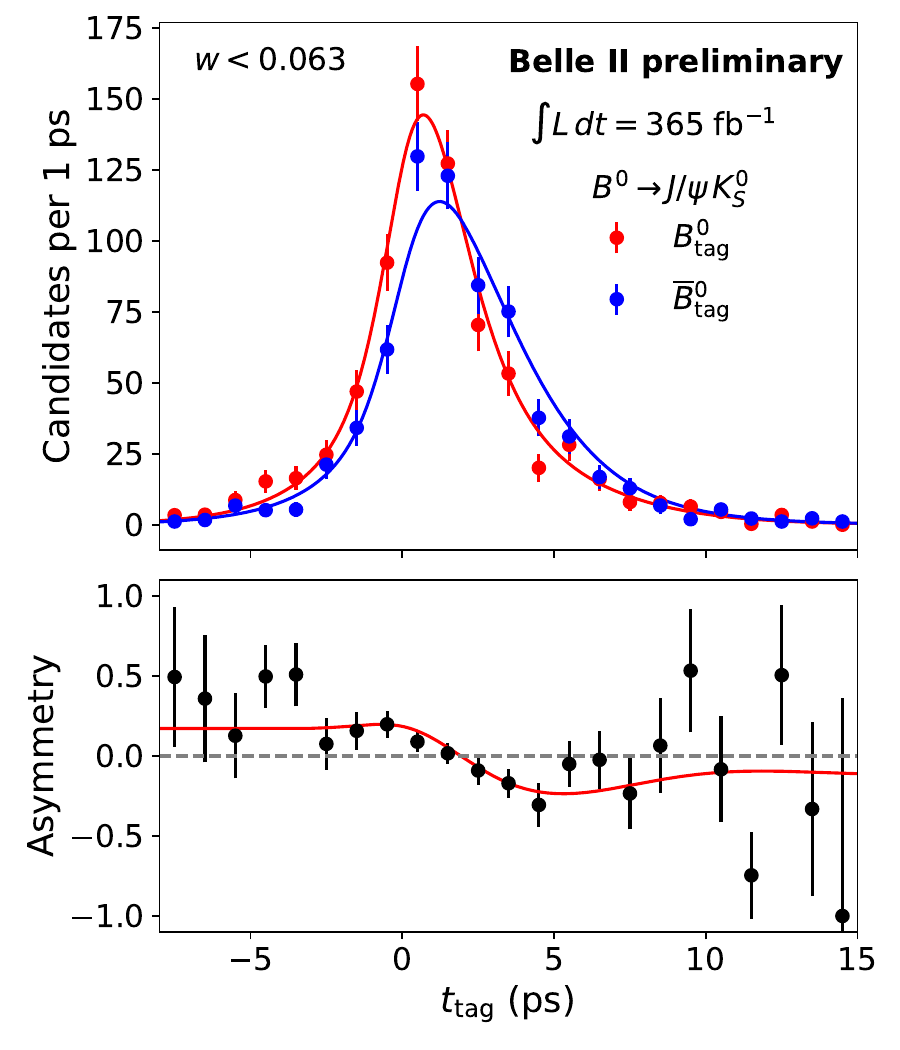}
\caption{(Left) Distribution of the tag-$B$ decay time $t_{\rm tag}$ for background-subtracted $B^0 \to D^-\pi^+$ decays, accompanied by a (red) opposite-flavour and (blue) same-flavour tag $B$ meson, with flavour-mixing fit projections overlaid. (Right) Distribution of the tag-$B$  decay time $t_{\rm tag}$ for background-subtracted $B^0 \to J/\psi K^0_S$ decays, accompanied by a (blue) $\overline{B}{}^0$ or (red) $B^0$ tag meson with projections of decay-time dependent \textit{CP}-violation fit overlaid. Both distributions are restricted to decays with highest flavour-tagging quality (about 24\% of the signal yield) to aid visualization. Bottom panels show the corresponding yield asymmetries as functions of tag-$B$ decay time; asymmetries are normalized differences between yields of (left) direct decays and decays following an oscillation and (right) signal decays accompanied by a $B^0$ and  $\overline{B}{}{^0}$ tag meson.}
\label{fig:fitProjections-control}
\end{figure*}

We validate the fit model with multiple closure tests. Fits to a realistic simulated sample four times larger than the data yield unbiased results, and ensembles of simulated experiments confirm unbiased estimators with correct uncertainty coverage. The continuum parameterisation is further validated by fitting events in signal-depleted $\Delta E$ and $M_{\rm bc}$ sidebands in data.

The $t_{\rm tag}$-resolution function is developed in simulation from the distribution of the difference
between reconstructed and true $t_{\rm tag}$. It is parameterised by a Gaussian core with asymmetric tails,
with width and tail parameters free to vary smoothly as functions of the per-event uncertainty measurement
$\sigma_{t_{\rm tag}}$. The resolution function is then calibrated in data using 36\,000 
\mbox{$B^{0}\to D^{-}\pi^{+}$} decays.  Measurements of the $B^0$ lifetime and $B^{0}$--$\overline{B}{}^0$ mixing frequency in this sample provide reliable benchmarks of the time-evolution model.
 After a kinematic selection, we statistically subtract residual backgrounds using the so-called $s$Weight technique~\cite{PIVK2005356}, based on the results of a fit to the $\Delta E$ distribution. We adjust the resolution parameters by fitting the background-subtracted $t_{\rm tag}$ distribution of the $B^{0}\to D^{-}\pi^{+}$ decays (Fig.~\ref{fig:fitProjections-control} (left)).  We measure the $B^0$ lifetime 
$\tau = 1.524 \pm 0.026\,\mathrm{ps}$ and the $B^{0}$--$\overline{B}{}^0$ mixing frequency $\Delta m_d = 0.512 \pm 0.011\,\mathrm{ps}^{-1}$, consistent with the known values~\cite{ParticleDataGroup}. After the validation, the resolution-function parameters are determined by a fit with lifetime and mixing-frequency fixed to the known values to further improve model accuracy. 
The resulting resolution function is used in the $B^{0}\to\pi^{0}\pi^{0}$ PDF, and a systematic uncertainty is assigned to cover possible modelling differences between the control and signal topologies.

To validate the $t_{\rm tag}$-based approach, we measure time-dependent \textit{CP}~violation  without using the signal vertex in the decay $B^{0}\to J/\psi K^{0}_{\rm S}$, which provides a low-background, high-precision benchmark. We apply the same selection and sequential fit strategy used in the conventional $\Delta t$-based analysis of Ref.~\cite{flavorTagger}: a fit to the $\Delta E$ distribution determines the sample composition
and provides $s$Weights for background subtraction, followed by a fit to the $(q,w,t_{\rm tag},\,\sigma_{t_{\rm tag}})$ distributions (Fig.~\ref{fig:fitProjections-control} (right)) to extract the mixing-induced and direct \textit{CP}-violating parameters $S_{J/\psi K^{0}_{\rm S}}$ and $C_{J/\psi K^{0}_{\rm S}}$. From about 6\,900 $B^{0}\to J/\psi K^{0}_{\rm S}$ decays, and using the calibrated $t_{\rm tag}$ resolution, we measure \mbox{$S_{J/\psi K^{0}_{\rm S}}= 0.835 \pm 0.104$} and \mbox{$C_{J/\psi K^{0}_{\rm S}}= -0.056 \pm 0.033$}, consistent with both the known values and  those obtained with the conventional $\Delta t$-based analysis on the same data~\cite{flavorTagger}.

\begin{figure*}[tb]
\centering
\includegraphics[width=120pt]{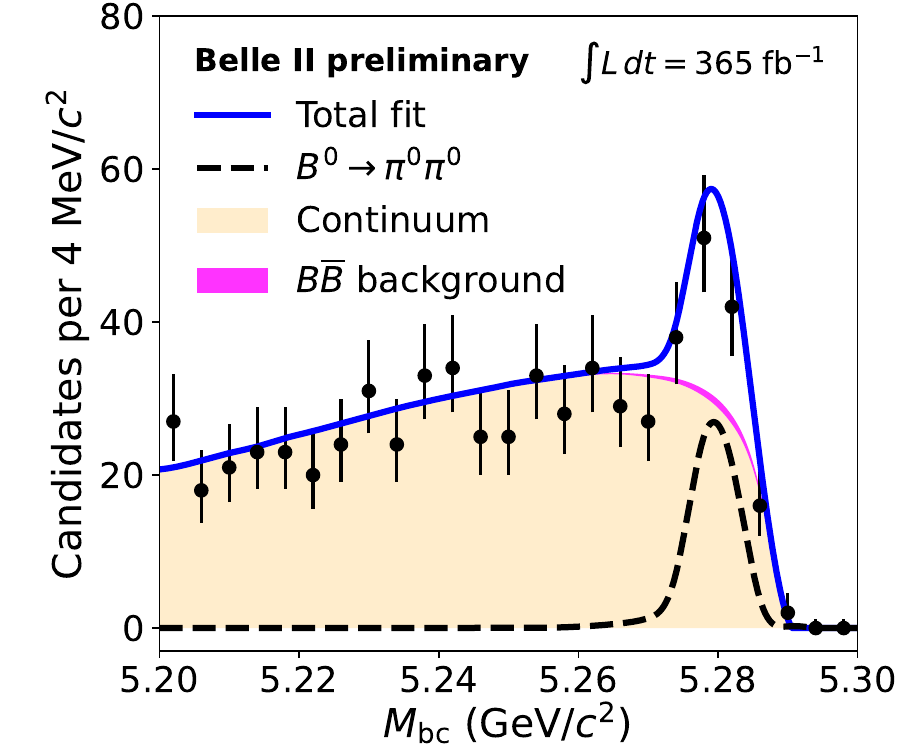}
\includegraphics[width=120pt]{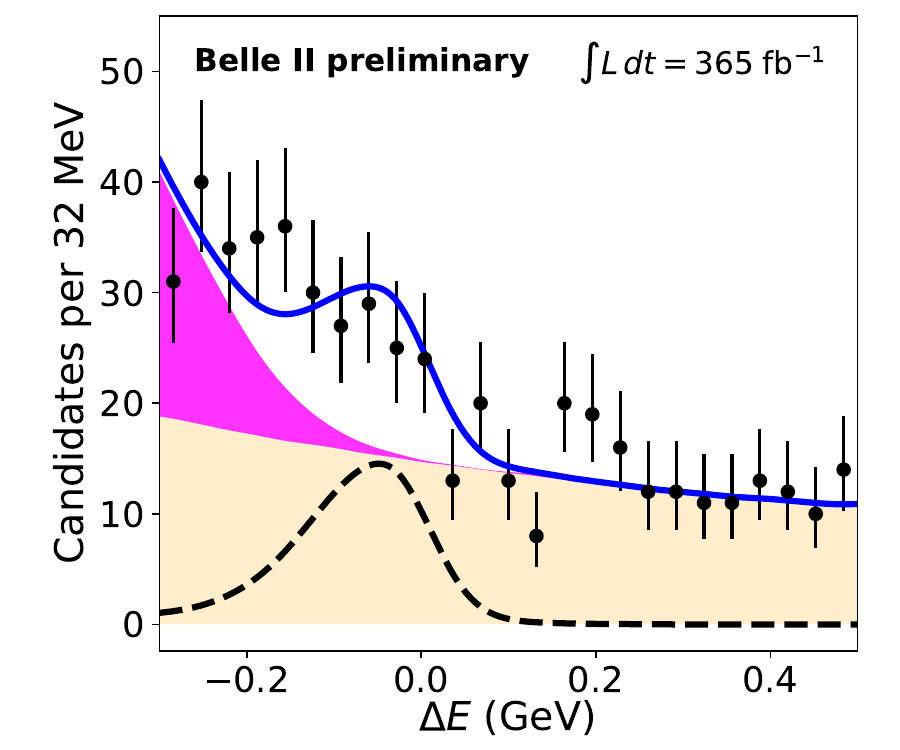}
\includegraphics[width=120pt]{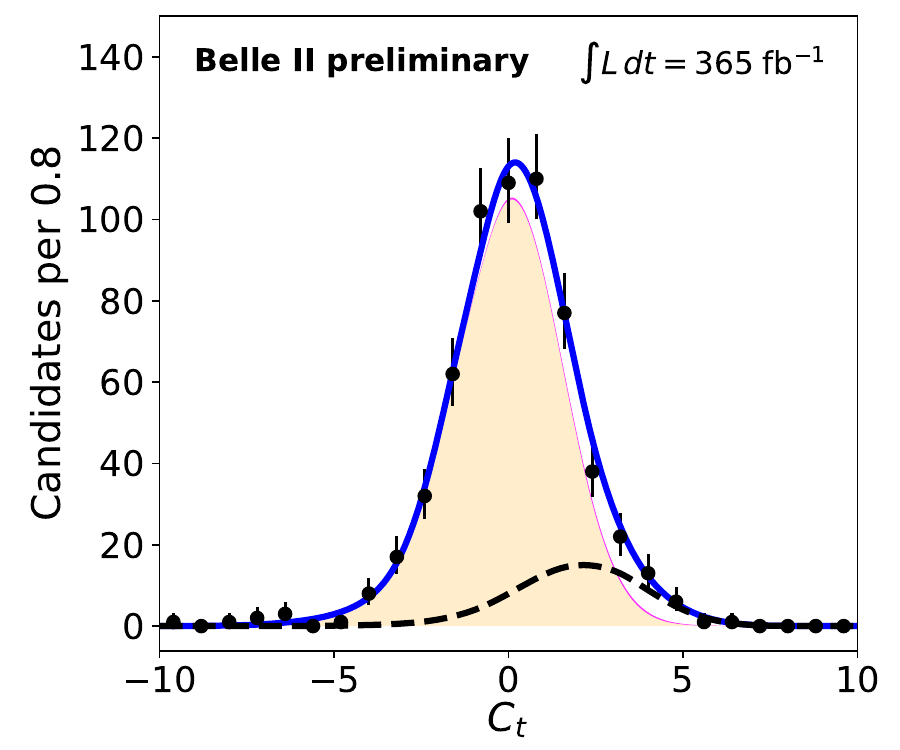}
\includegraphics[width=120pt]{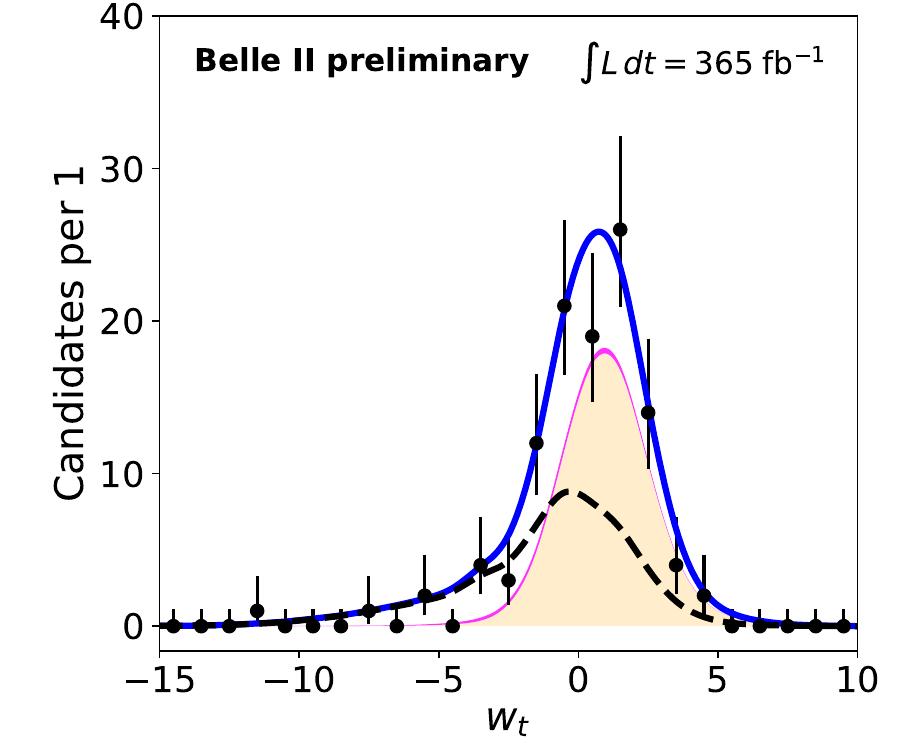}
\includegraphics[width=120pt]{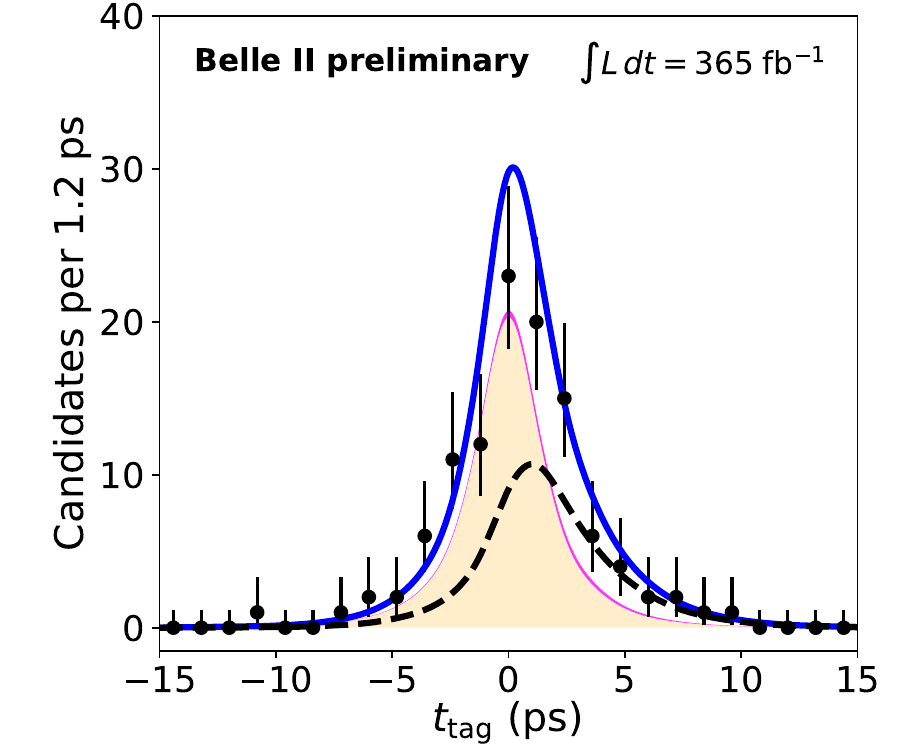}
\includegraphics[width=120pt]{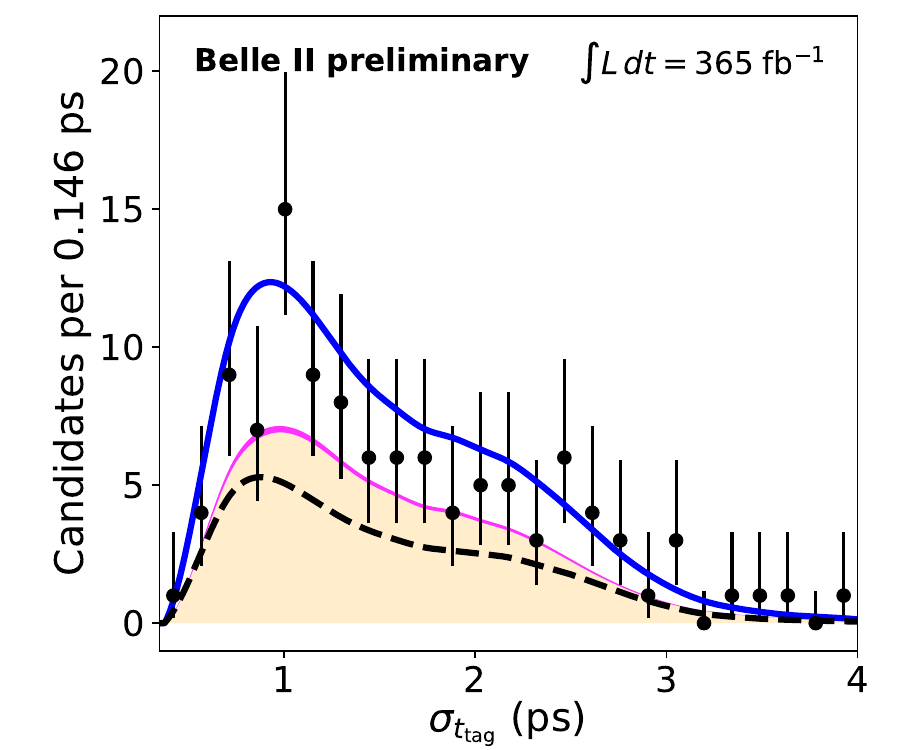}
\caption{Distributions of discriminating observables for signal-enhanced  $B^0 \to \pi^0\pi^0$ candidates reconstructed in data with fit projections overlaid: (top to bottom, left to right) beam-constrained invariant mass $M_{\rm bc}$; $B$-energy difference $\Delta E$; transformed continuum-suppression output $C_t$; transformed wrong-tag fraction $w_t$; tag-$B$ decay time $t_{\rm tag}$; and tag-$B$ decay-time uncertainty $\sigma_{t_{\rm tag}}$. The sample components are illustrated in the legend. 
The signal-enhancing selections are  $5.275 < M_\mathrm{bc} < 5.285$\,GeV/$c^2$, $-0.1 < \Delta E < 0.05$\,GeV,  and $C_t > 1.8$ (with each requirement omitted when plotting the corresponding distribution).} 
\label{fig:fitProjections_sigenh}
\end{figure*}

\begin{figure}[tb]
\centering
\includegraphics[width=160pt]{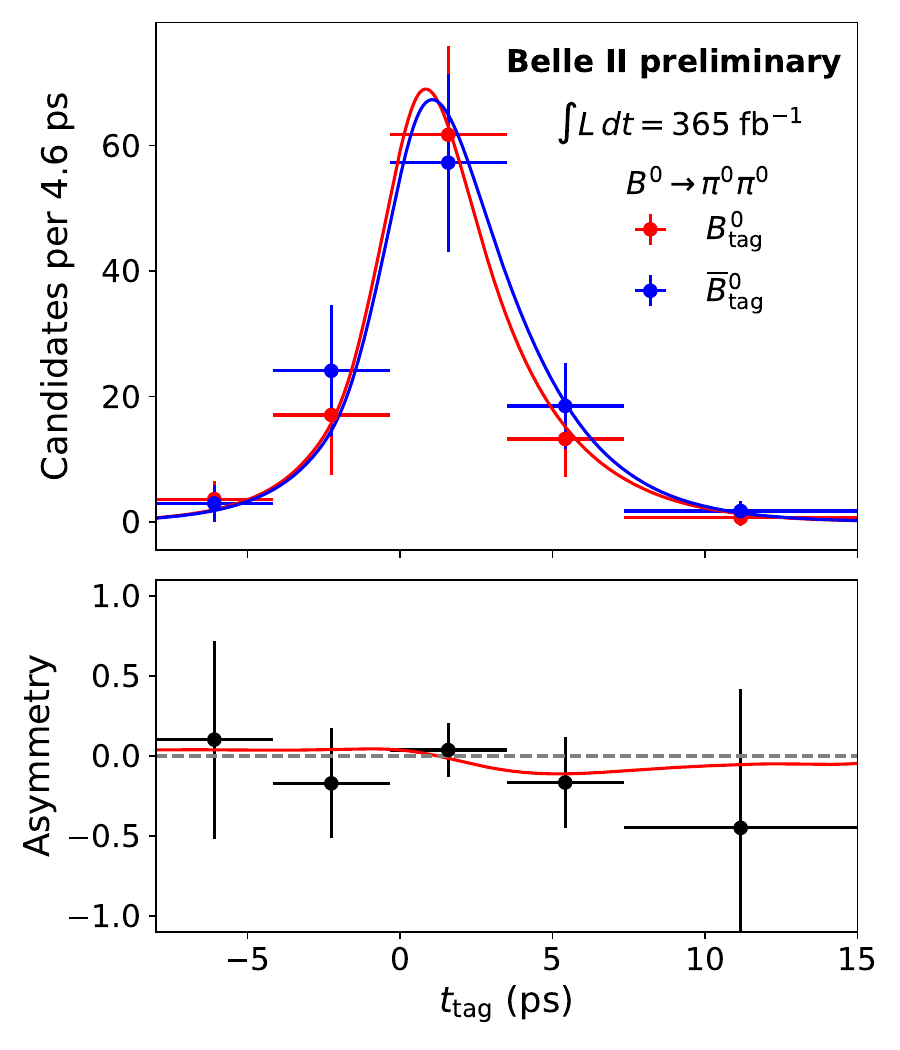}
\caption{Distribution of the tag-$B$ decay time $t_{\rm tag}$ for background-subtracted $B^0 \to \pi^0\pi^0$ decays,  accompanied by a (blue) $\overline{B}{}^0$ or (red) $B^0$ tag meson with fit projections overlaid. The background is subtracted using  $s$Weigths~\cite{PIVK2005356}. The bottom panel shows the corresponding yield asymmetry as a function of tag-$B$  decay time. The asymmetry is the normalized difference between yields of signal decays accompanied by a $B^0$ and $\overline{B}{}{^0}$ tag meson.}
\label{fig:fitProjections_ttag_signal}
\end{figure}

We then turn to our signal $B^0\to \pi^0\pi^0$ sample.
 Fit projections in a signal-enhanced region are shown in Fig.~\ref{fig:fitProjections_sigenh}.  The distribution of $t_{\rm tag}$ separated for $B_{\rm tag}$ candidates tagged as $B^0$ or $\overline{B}{}{^0}$, and the resulting time-dependent asymmetry, are reported in Fig.~\ref{fig:fitProjections_ttag_signal}.  
The fit yields  $22\,320   \pm 150\,(\mathrm{stat})$ continuum events,  $203 \pm 27\,(\mathrm{stat})$ $B\overline{B}$ background events, and $171 \pm 24\,(\mathrm{stat})$ signal $B^0\to \pi^0\pi^0$  decays, with time-dependent \textit{CP}-violating parameters 
\begin{align}
S_{\pi^0\pi^0} &= 0.61^{+0.75}_{-0.79}\,(\mathrm{stat}) \pm 0.11\,(\mathrm{syst})\,, \\
C_{\pi^0\pi^0} &= 0.05 \pm 0.28\,(\mathrm{stat})\pm 0.07\,(\mathrm{syst})\,. 
\end{align}
The linear correlation between the parameters is 6.6\%.  Uncertainties are dominated by sample size. Systematic uncertainties are assessed by varying the fit model and its calibration inputs; by ensemble tests of estimator properties; and by accounting for uncertainties in the flavour-tagging calibration, the $t_{\rm tag}$ resolution model, and possible \textit{CP}-asymmetry effects in the $B\overline{B}$ background. The contributions from the individual sources are reported in  Methods~Tab.~\ref{table:uncertainty}; the largest is from the $t_{\rm tag}$ resolution model. 
The $C_{\pi^0\pi^0}$ result agrees with the time-integrated asymmetry reported in Ref.~\cite{BelleIIpi0pi0BR}, is more precise and supersedes it. The $S_{\pi^0\pi^0}$ result is the first ever determination of mixing-induced {\it CP}-violation in $B^0\to \pi^0\pi^0$ decays.

\begin{figure*}[tb]
\centering
    \includegraphics[width=0.7\linewidth]{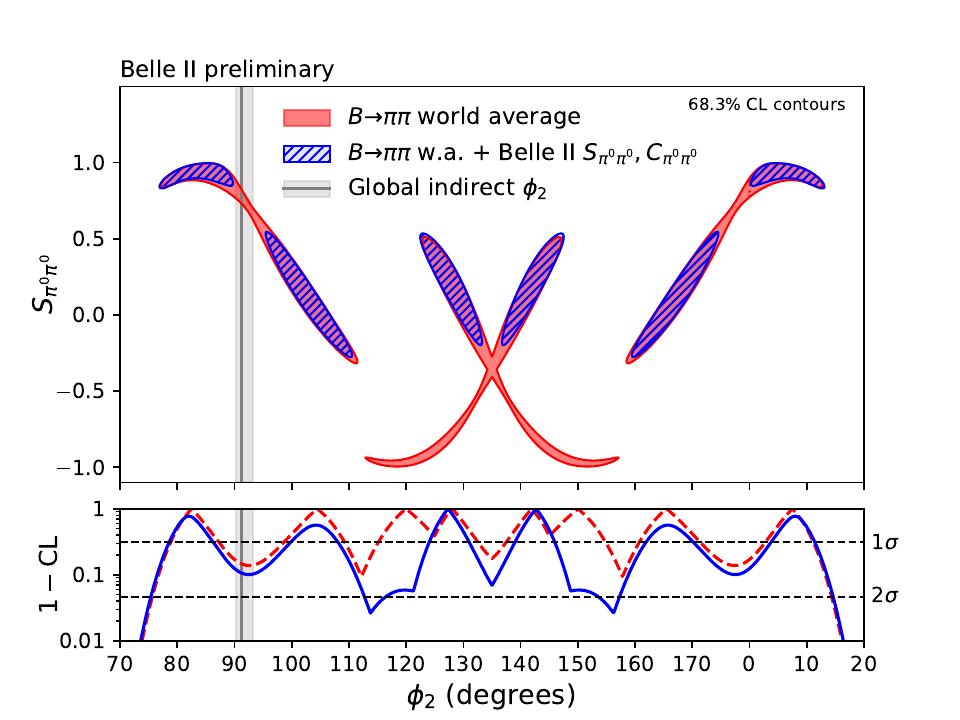}
    \caption{Constraints on $\phi_2$ from the $B\to\pi\pi$ isospin analysis. The upper panel shows the $68.3\%$ CL allowed regions in the $(\phi_2,S_{\pi^0\pi^0})$ plane; the lower panel shows the corresponding one-dimensional $(1-\mathrm{CL})$  curves as a function of $\phi_2$. In both panels, red regions and curves show the result obtained without the $S_{\pi^0\pi^0}$ and $C_{\pi^0\pi^0}$ measurements reported in this work, while blue regions and curves show the result obtained when they are included. The vertical grey band indicates the global constraint on $\phi_2$ from indirect measurements, assuming the Standard Model~\cite{Charles:2004jd}. 
    The horizontal axis reflects the $180^\circ$  periodicity of $\phi_2$.}
    \label{fig:alpha_vs_S00}
\end{figure*}

We translate the measured  values of $S_{\pi^0\pi^0}$ and $C_{\pi^0\pi^0}$ into constraints on the \textit{CP}-violating phase $\phi_{2}$. We use the standard isospin analysis of $B\to\pi\pi$ amplitudes~\cite{Gronau:1990ka,Charles:2017evz}, adding our $S_{\pi^0\pi^0}$ and $C_{\pi^0\pi^0}$ measurements to the other available \mbox{$B\to\pi\pi$} inputs (see Methods~Tab.~\ref{tab:alphaInputs}).  
Figure~\ref{fig:alpha_vs_S00} shows the 68.3\% confidence-level (CL) region in the $(\phi_2,S_{\pi^0\pi^0})$ plane and the value of $(1 - \rm{CL})$ as a function of $\phi_2$. Our results, and in particular the inclusion of $S_{\pi^0\pi^0}$  in the isospin analysis, reduce the number of discrete solutions from eight to six, removing the 68.3\% CL region for \mbox{$S_{\pi^0\pi^0} < -0.25$}. The updated value of $C_{\pi^0\pi^0}$ also slightly shifts the peaks in the $(1 - \rm{CL})$ curve as a function of $\phi_2$. Overall, our measurement shortens  the allowed  68.3\% CL $\phi_2$ interval by 40\%.  
The solution preferred by the Standard-Model constraints is $(82.0^{+4.4}_{-3.5})^\circ$, in agreement with the determination from the $\rho\rho$ system and similar in precision~\cite{Belle-II:2024frs}. 

In summary, we report the first measurement of mixing-induced \textit{CP} violation in the  decay
\mbox{$B^{0}\!\to\!\pi^{0}\pi^{0}$}.
The result is enabled by a novel method that, exploiting the quantum entanglement of the $B^{0}\overline{B}{}^0$ pair, infers the time evolution of the system from the tag-$B$ decay time
$t_{\rm tag}$ without signal $B$ vertex reconstruction. The analysis relies on a data-driven calibration of the $t_{\rm tag}$ resolution using
$B^{0}\to D^{-}\pi^{+}$ decays and is validated with $B^{0}\to J/\psi K^{0}_{\rm S}$ decays,
demonstrating that the $t_{\rm tag}$-based approach reliably reproduces  established time-dependent
\textit{CP}-violation results.
Using $190$ million $\Upsilon(4S)\to B^{0}\overline{B}{}^0$ decays collected with Belle~II, we measure $S_{\pi^{0}\pi^{0}}$ with a precision that would require a data set twenty times larger with a conventional time-dependent analysis. We also measure   $C_{\pi^{0}\pi^{0}}$, superseding the previous Belle~II result~\cite{BelleIIpi0pi0BR}.
Incorporating these results into the \mbox{$B\to\pi\pi$} isospin analysis reduces the
discrete ambiguities and strengthens the constraint on the \textit{CP}-violating phase $\phi_{2}$  from $B \to \pi\pi$ decays. In a broader perspective, this result shows that the $t_{\rm tag}$-based approach makes \mbox{$B \to \pi\pi$} decays competitive in the determination of $\phi_2$ much earlier than previously anticipated, thereby enabling more incisive comparisons among complementary channels, and accelerating progress on $\phi_2$.
With larger Belle~II data sets, the  $t_{\rm tag}$-based method provides a path to more precise tests of the CKM framework in decay channels without reconstructible decay vertices.

\clearpage
\newpage

\section*{Methods}\label{sec11}

\subsection*{Experimental setup}
The data were collected with the Belle~II detector at the SuperKEKB collider in Tsukuba, Japan~\cite{AKAI2018188,abe2010belleiitechnicaldesign}.  
SuperKEKB is an electron--positron collider whose main rings store a 7.0\,GeV electron beam and a 4.0\,GeV positron beam, supported by a linear injector and a positron damping ring. 
It employs the nanobeam collision scheme: low-emittance beams are strongly focused by final-focus superconducting quadrupole magnets and collide with a horizontal crossing angle of $83\,\mathrm{mrad}$. 
This reduces the longitudinal overlap of the colliding bunches, mitigates the hourglass effect---the luminosity loss caused by collisions occurring outside the narrowest part of the focused beams---and produces a luminous region with characteristic dimensions of about $13\,\mu\mathrm{m}$, $0.2\,\mu\mathrm{m}$, and $350\,\mu\mathrm{m}$ in $x$, $y$, and $z$, respectively. 
The coordinate system is defined with the $z$ axis aligned to the symmetry axis of the Belle~II solenoid, which points approximately along the direction of the electron-beam;  
the $x$ axis in the horizontal plane pointing outside the ring; and the $y$ axis pointing vertically upward. 
The compact luminous region and crossing geometry are crucial for the present measurement because they constrain the production point of the entangled $B^0\overline{B}{}^0$ pair, enabling the determination of $t_{\rm tag}$.

Belle~II is a large-solid-angle magnetic spectrometer surrounding the interaction region. The determination of $t_{\rm tag}$ relies mostly  on a two-layer pixel detector~\cite{Belle-IIDEPFET:2021pib}, which provides precise vertex measurements close to the beam pipe. Of the outer layer, only one sixth is installed for the data used in this work. The  tag-$B$ vertex resolution is about $30\,\mu\mathrm{m}$ in all directions. The reconstruction of $B^0 \to \pi^0\pi^0$ decays relies on the electromagnetic calorimeter~\cite{Miyabayashi_2020}, which is made of 8736 CsI(Tl) crystals and provides efficient photon reconstruction, with a mass resolution of about $8$\,MeV for neutral pions from $B^0 \to \pi^0\pi^0$ decays.

\subsection*{Signal selection}
We select signal candidates with an online event filter followed by offline reconstruction. The online selection requires events to satisfy criteria based on total energy and charged-particle multiplicity to preferentially retain collisions that produce hadrons, and is fully efficient for the signal.

In the offline analysis, we identify photon candidates as energy deposits in the calorimeter (clusters) larger than 30 MeV and that involve more than one crystal, to reject noise, which is uncorrelated across calorimeter channels. 
We require the cluster-detection time to be within 200\,ns of the collision time, to suppress energy deposits from secondary hadronic interactions unrelated to the relevant collision and overlapping beam-induced photons. Multiplicative photon-energy corrections ranging from 0.990 to 1.010, with 0.15\%--0.50\% uncertainties, are derived from control samples in data and used to correct the absolute calorimeter-energy scale. 

\begin{figure*}[tb]
    \centering{
    \includegraphics[width=180pt]{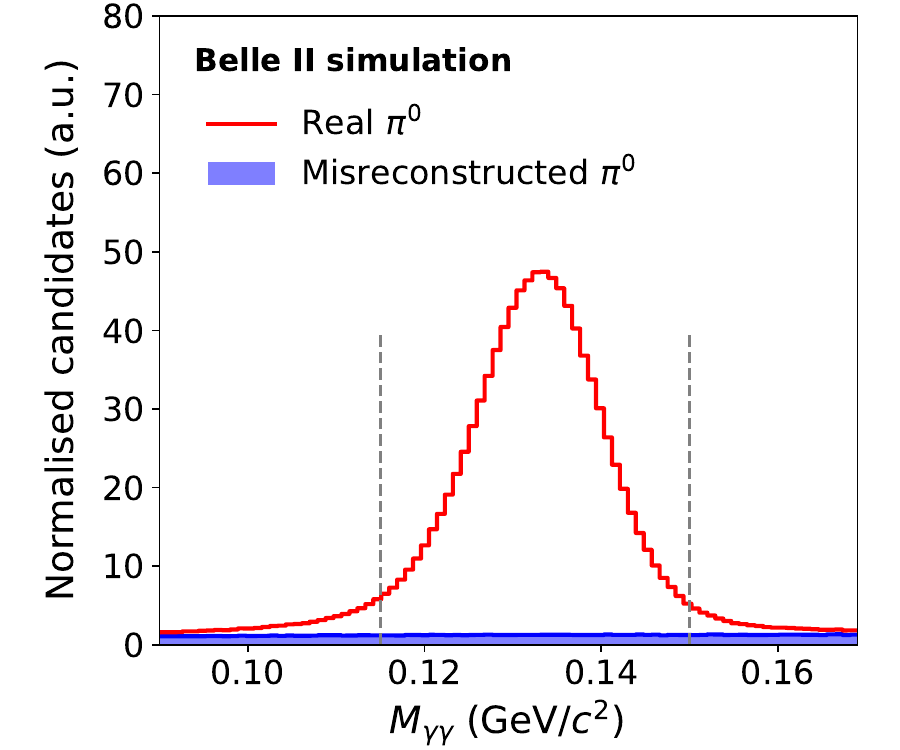} \hspace{1.5cm}
    \includegraphics[width=180pt]{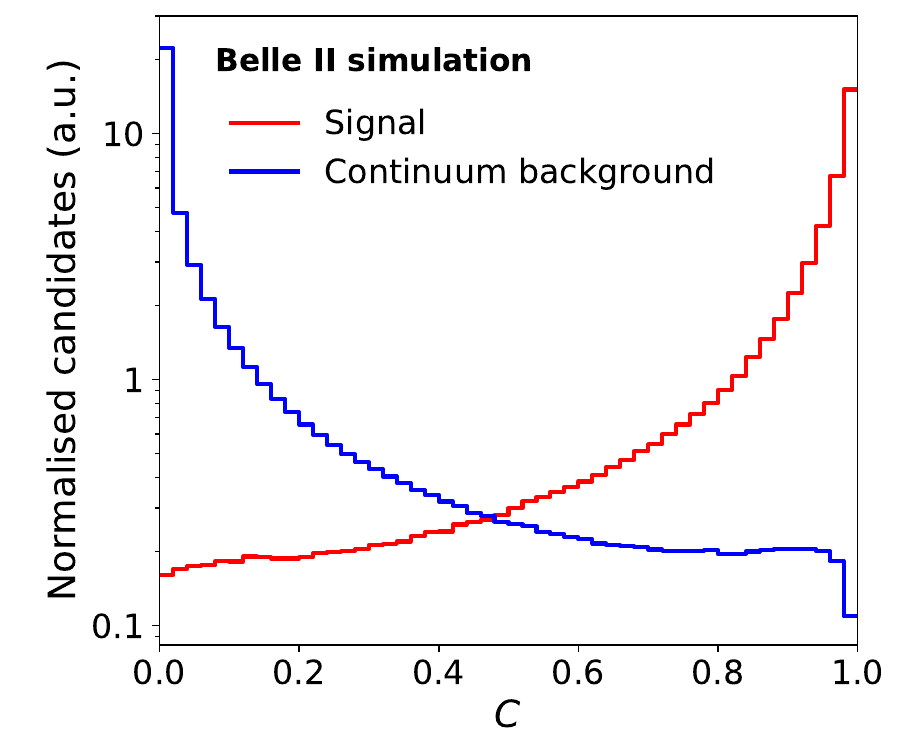}
    }
    \caption{(Left) Distribution of the diphoton mass $M_{\gamma\gamma}$ in simulation after all photon and $\pi^0$  selections except the diphoton-mass requirement. The dashed vertical lines indicate the signal-region boundaries. The properly reconstructed $\pi^0$ component is shown in red, stacked on the misreconstructed component shown in blue. (Right) Distribution of the output $C$ of the continuum-suppression classifier for simulated (blue) continuum and (red) signal $B^0 \to \pi^0 \pi^0$ events.}
    \label{fig:selection}
\end{figure*}

The high momentum of signal $\pi^0$'s offers a natural suppression of misreconstructed and beam-induced photons. To reduce contributions from these sources further, we use a boosted decision  tree~\cite{Keck:2017gsv} trained to discriminate simulated signal photons from beam-induced and misreconstructed photons using nine discriminating observables: the energy detected in the crystal with the largest signal, three observables that describe the energy sharing among crystals, the photon's momentum transverse to the beam direction, the distance between the cluster and the trajectory of the nearest charged particle, the number of crystals in the cluster, the polar angle coordinate of the cluster, and one observable that describes the fraction of cluster energy detected in the central crystal. To reproduce realistic experimental conditions, beam-induced photons from data are overlaid on simulated events.
We choose the threshold on the boosted-decision-tree output that maximizes the yield of signal photons over the square root of the sum of misreconstructed and beam-induced photon yields, as expected from simulation. 
The selection removes 83\% of misreconstructed and beam-induced photons and retains 96\% of signal photons.

We form $\pi^0$ candidates by pairing the selected photon candidates. To suppress combinatorial background from low-energy photons, we require that the $\pi^0$ momentum be greater than 1.5\,GeV/$c$, and the angle between the momenta of the photon candidates be less than 0.4 radians. In the $\pi^0$ candidate rest frame, the magnitude of the cosine of the angle between the photon-candidate direction and the boost direction from the laboratory frame must not exceed  0.98 to suppress misreconstructed $\pi^0$ mesons, which tend to peak near~1.00. The diphoton mass is required to lie between 0.115 and 0.150\,GeV/$c^2$, corresponding to approximately $-2.5$ and $+2.0$ units of resolution around the known $\pi^0$ mass---see  Fig.~\ref{fig:selection}~(left).  The asymmetric range compensates for calorimeter energy leakage. We apply a kinematic fit that improves momentum resolution by constraining the diphoton mass to the
known $\pi^0$ mass. The photon and $\pi^0$ reconstruction performance is  validated using  \mbox{$D^{*+} \to D^0(\to K^-\pi^+\pi^0)\pi^+$} decays reconstructed in data.
  
Finally, we train a boosted-decision-tree classifier $C$ to discriminate signal from continuum by analysing 27 observables, comprising modified Fox-Wolfram moments~\cite{FoxWolfram1978, Abe2001PLB, Abe2001PRL}, sphericity-related quantities~\cite{PhysRevD.1.1416}, thrust-related quantities~\cite{Farhi:1977sg}, and energy detected in sets of concentric cones with various opening angles centred around the thrust axis of the event~\cite{Brandt:1964sa}. The output of this classifier is shown in  Fig.~\ref{fig:selection}~(right), comparing signal and background from simulation; the threshold that optimises sensitivity to $S_{\pi^0\pi^0}$ is 0.7.  Compared to the time-integrated analysis~\cite{BelleIIpi0pi0BR}, this different continuum-suppression selection increases the signal efficiency by 30\%, and it improves the expected statistical precision on $S_{\pi^0\pi^0}$ by 15\%, even though it triples the continuum yield.
We~further restrict the sample by requiring candidate signal mesons to have \mbox{$M_{\rm bc} > $ 5.2~{\rm GeV}/$c^2$} and \mbox{$-0.3  < \Delta E < 0.5~{\rm GeV}$}.

The $t_{\rm tag}$ decay time is defined as
\begin{equation}
\label{eqn_ttag_def}
    t_{\rm tag} = m_{B} \frac{\vec{L} \cdot \vec{p}_{\rm tag} }{|p_{\rm tag}|^2}\,,
\end{equation}
in which $m_{B}$ is the known $B^0$ mass, $\vec{L}$ is the flight-distance vector that connects the production and decay space-points of the tag $B$ meson, and $\vec{p}_{\rm tag}$ is its momentum vector. The latter is  inferred as $\vec{p}_{\rm tag} = \vec{p}_{\Upsilon(4S)} - \vec{p}_{\pi^0\pi^0}$, where $\vec{p}_{\Upsilon(4S)}$ is the momentum of the $\Upsilon(4S)$ meson in the \mbox{$e^+e^-\to\Upsilon(4S)\to B^0 \overline{B}{}{^0}$} process, which is known precisely from the accelerator parameters, and $\vec{p}_{\pi^0\pi^0}$ is the signal momentum vector. 
We reconstruct $t_{\rm tag}$ for each event by implementing the fit of Eqs.~(11--12) of Ref.~\cite{rzr8-l6l8}, which also returns the uncertainty  $\sigma_{t_{\rm tag}}$. 
A requirement $\sigma_{t_{\rm tag}} > 0.35\,\mathrm{ps}$ is also imposed; it rejects 8\% of continuum, while being fully efficient for signal.  
We also require $\sigma_{t_{\rm tag}} < 4$\,ps and $|t_{\rm tag}|< 15$\,ps as fiducial requirements to avoid candidates with a misreconstructed decay time.

\subsection*{Flavour tagging} 
We infer the tag-$B$ flavour  using a graph-neural-network algorithm~\cite{flavorTagger}, which provides a flavour decision $q=+1$ ($-1$) for a $B^0$ ($\overline{B}{}^0$) tag and an event-by-event estimate of a wrong-tag probability $w$. The tagging efficiency $\epsilon$, defined as the fraction of events assigned a flavour decision, is close to unity. 

\begin{figure}[tb]
 \centering
  \includegraphics[width=180pt]{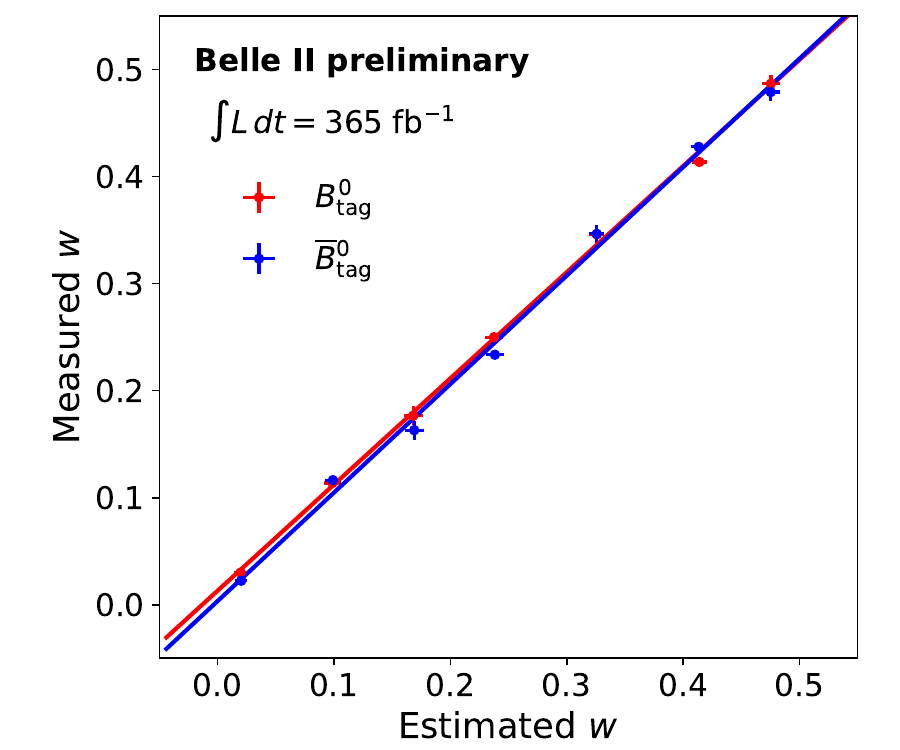}
 \caption{Measured wrong-tag fractions in data using \mbox{$B^0 \to D^{(*)-}\pi^+$} decays as functions of the wrong-tag fractions estimated by the flavour-tagging algorithm for (red) $B^0$- and (blue) $\overline{B}{}{^0}$-tag mesons with fit projections overlaid.}
 \label{fig:calibGFlaT}
\end{figure}

The wrong-tag probability predicted by the tagger is calibrated using low-background control samples of flavour-specific $B^0 \to D^{(*)-}\pi^+$ decays~\cite{flavorTagger}, and its average is found to be about~23\%.  
A linear correction is applied separately for $B^0_{\rm tag}$ and $\overline{B}{}^0_{\rm tag}$ candidates to account for small differences between the predicted and observed mistag probabilities in data. The corresponding calibration functions are shown in  Fig.~\ref{fig:calibGFlaT}. Possible differences in tagging efficiency between $B^0_{\rm tag}$ and $\overline{B}{}^0_{\rm tag}$ mesons are described by an asymmetry parameter $a_{\rm tag}$,  
which is found to be consistent with zero in the same control channels. We verify with simulated events  that the calibration derived from the control modes is also valid for the signal decay.

The signal time-dependent PDF of Eq.~(\ref{eq:prob}) is expanded to include flavour-tagging imperfections, as 
\begin{equation}
\label{eq:prob_ft}
\begin{split}
\mathcal{P}(t_{\rm tag},q)= 
&\frac{e^{-t_{\rm tag}/\tau}}{2\tau}
\biggl(
1 + q\bigl[a_{\rm tag}(1-2\overline{w})-\Delta w\bigr] \\
&- q\bigl[1-2\overline{w}+qa_{\rm tag}-a_{\rm tag}\Delta w\bigr]  \\
& \times D \Big[ S_{\pi^0\pi^0}\,\sin [\Delta m (t_\mathrm{tag}-\hat{t})] \\
&+ C_{\pi^0\pi^0} \cos[ \Delta m (t_\mathrm{tag}-\hat{t})] \Big]  \biggr).
\end{split}
\end{equation}
Here $\overline{w}=(\omega_B+\omega_{\overline{B}})/2$ is the average of the mistag fractions for true flavours $\omega_B$ and $\omega_{\overline{B}}$, and \mbox{$\Delta w=\omega_B-\omega_{\overline{B}}$} is their difference. We express $\overline{w}$ and $\Delta w$ in terms of the per-event mistag estimate $w$ provided by the algorithm using the linear functions obtained in the flavour-tagger calibration.

\subsection*{Tag-$B$ decay-time resolution}

\begin{figure}[tb]
 \centering
  \includegraphics[width=180pt]{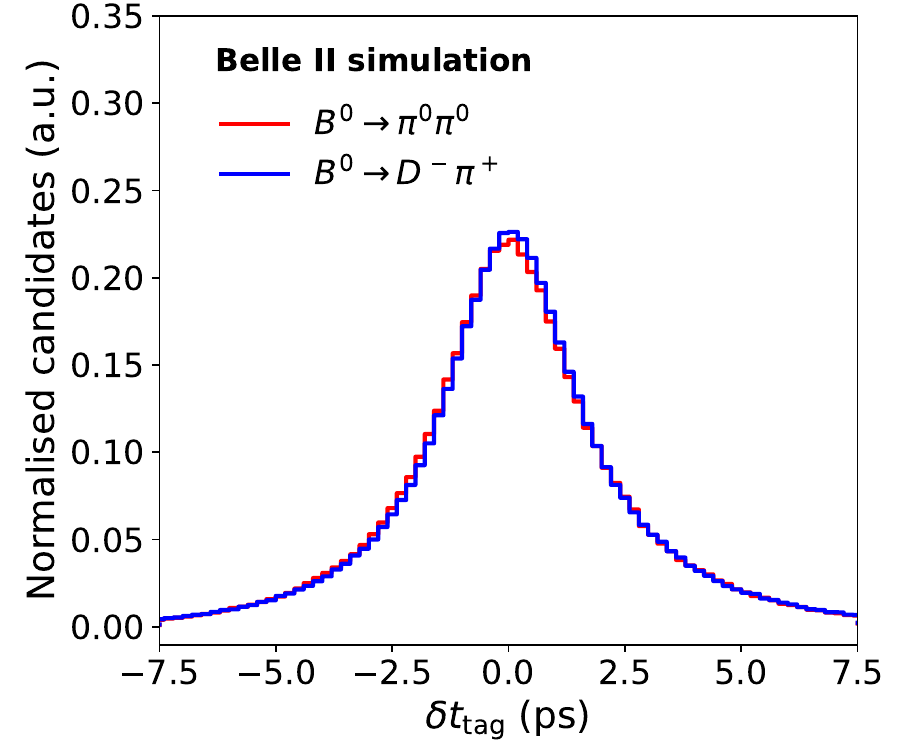}
 \caption{Distribution of the difference between the measured and true values of $t_{\rm tag}$,  obtained in simulation for (blue) $B^0 \to D^-\pi^+$ and (red) $B^0 \to \pi^0\pi^0$ decays. The distribution for $B^0 \to J/\psi K_{\rm S}^0$ overlaps that for $B^0 \to D^-\pi^+$ and is not shown. 
}
 \label{fig:resolutionFunction_Dpi_MC}
\end{figure}

The $t_{\rm tag}$ resolution function describes the difference between the measured and true values of $t_{\rm tag}$. It is  
parameterised in simulation as
\begin{equation}
\label{eq:res}
\begin{split}
\mathcal{R}(t_{\rm tag}) =\,
& f_1\, G_1(t_{\rm tag}\mid \mu_1,\sigma_1) \\
+\,
&(1-f_1)\,
\Big[
\exp_{LR}(t_{\rm tag}\mid \tau_L,\tau_R) \\
&\qquad \qquad \otimes
G_2(t_{\rm tag}\mid \mu_2,\sigma_2)
\Big],
\end{split}
\end{equation}
where $G_1$ ($i=1,2$) are Gaussian functions, with means $\mu_i$, standard deviations $\sigma_i$; $\exp_{LR}$ is an asymmetric two-sided exponential distribution with slopes $\tau_L$ and $\tau_R$; and $f_1$ is the fractional area of the core Gaussian function. The width and slope parameters, as well as the core fraction, are allowed to vary smoothly with~$\sigma_{t_{\rm tag}}$. 
 Figure~\ref{fig:resolutionFunction_Dpi_MC} shows 
the resolution function in simulation integrated over all values of $\sigma_{t_{\rm tag}}$.

\begin{figure}[tb]
 \centering
  \includegraphics[width=180pt]{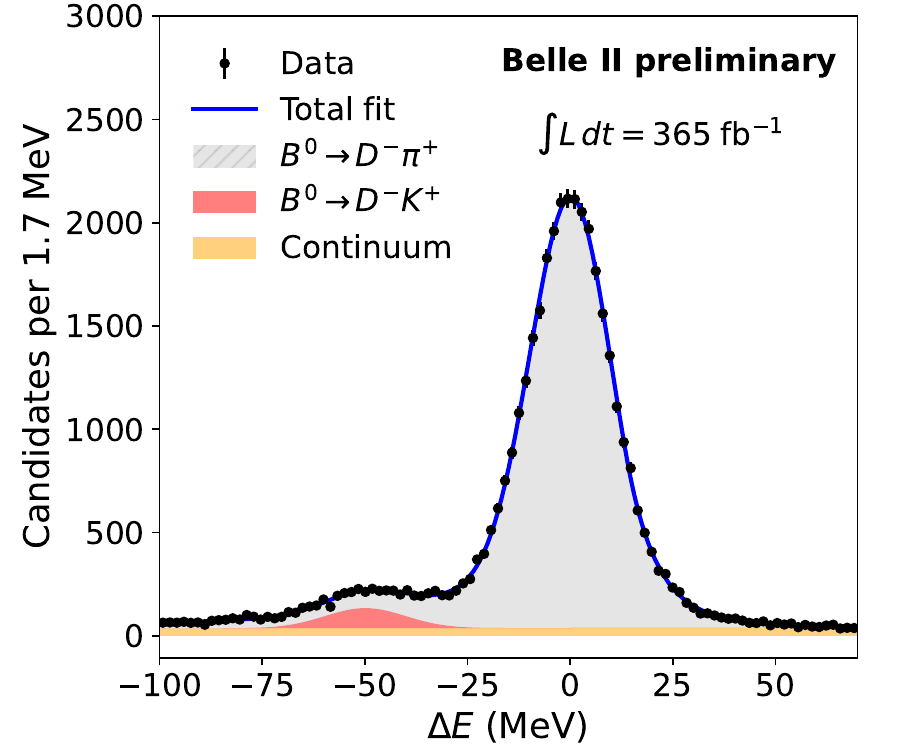}
 \caption{Distribution of $\Delta E$ for $B^0 \to D^-\pi^+$ candidates reconstructed in data, with fit projection overlaid.}
 \label{fig:deltaE_Dpi}
\end{figure}

We adjust the resolution function to data using $B^0 \to D^-(\to K^+\pi^-\pi^-)\pi^+$ decays, through a measurement of the $B^0$ lifetime and the \mbox{$B^0$--$\overline{B}{}{^0}$} oscillation frequency. The control decays are reconstructed following Ref.~\cite{flavorTagger}, with a more restrictive selection on $M_{\rm bc}$ and $\Delta E$, a requirement that the measured $D$-meson flight distance be non-negative, and a multivariate selection to suppress continuum. As in the $B^0 \to \pi^0\pi^0$ selection, we require $\sigma_{t_{\rm tag}} > 0.35\,\mathrm{ps}$ and $|t_{\rm tag}|< 15$\,ps. The $\Delta E$ distribution of $B^0 \to D^-\pi^+$ candidates is shown in  Fig.~\ref{fig:deltaE_Dpi}. From a fit to this  distribution, used to compute $s$Weights, we obtain about $36\,000$ $B^0 \to D^-\pi^+$ decays, with small residual contributions from continuum events and $B^0 \to D^-K^+$ decays.  

The parameters of the resolution model are determined  with a maximum-likelihood  flavour-oscillation fit to the $s$Weighted unbinned $(q,t_{\rm tag},\sigma_{t_{\rm tag}})$ distributions  (Fig.~\ref{fig:fitProjections-control} (left)) in bins of flavour-tagging quality $w$, using the corresponding tagging calibration parameters~\cite{flavorTagger}. The signal PDF is obtained  by replacing the usual $\Delta t$ dependence~\cite{flavorTagger} with the dependence on $t_{\rm tag}$ in the standard flavour-specific $B^0$--$\overline{B}{}^0$ mixing expression,
\begin{equation}
\begin{split}
&\mathcal{P}_{\rm mix}(t_{\rm tag},q)
\propto \\
&\qquad e^{-t_{\rm tag}/\tau}
\left(
1 - q\,h\,D\cos[\Delta m (t_{\rm tag}-\hat{t})]
\right),
\end{split}
\end{equation}
and convolving it with the resolution function of Eq.~(\ref{eq:res}). Here $h=\pm1$ is the charge of the pion from the $B^0 \to D^-\pi^+$ decay; we neglect the $\mathcal{O}(10^{-4})$ wrong-sign contribution~\cite{Belle:2010igk,BaBar:2008xvo,LHCb:2020zae}. We obtain  
\mbox{$\Delta m = 0.512 \pm 0.011(\mathrm{stat})\,\mathrm{ps}^{-1}$} and \mbox{$\tau = 1.509 \pm 0.026(\mathrm{stat})\,ps,$} consistent with the known values. The agreement validates the decay-time resolution model used in the signal fit. To further improve model accuracy, the final resolution-function parameters are determined by a fit with lifetime and mixing frequency fixed to the known values.

\subsection*{Fit modelling}
We determine $S_{\pi^0\pi^0}$ and $C_{\pi^0\pi^0}$ with an extended maximum-likelihood fit to the unbinned distributions of $\Delta E$, $M_{\rm bc}$, $C_t$, $q$, $w_t$, $t_{\rm tag}$ and $\sigma_{t_{\rm tag}}$. The sample is modelled as a mixture of signal, $B\overline{B}$ background, and continuum events, and the likelihood is
\begin{equation}
\begin{split}
\mathcal{L} \propto\, & e^{-(N_\mathrm{sig}+N_{B\overline{B}} + N_{q\overline{q}})}\\
& \times \prod_{i=1}^{M}
\left[
N_{\rm sig}\,\mathcal{P}_{\rm sig}^i +
N_{B\overline{B}}\,\mathcal{P}_{B\overline{B}}^i +
N_{q\bar{q}}\,\mathcal{P}_{\rm cont}^i
\right],
\end{split}
\end{equation}
where $N_j$ is the yield of the component $j$ determined by the fit,  $\mathcal{P}_j^i$ denotes the PDF for event $i$ under component $j$, and $M$ is the total number of events.

For each component, the PDF is factorised into a time-independent part and a time-dependent part. The time-independent part describes the observables $(\Delta E,\,M_{\rm bc},\,C_t)$. The corresponding PDFs for signal and $B\overline{B}$ background are determined from simulation, and expanded with additional, constrained degrees of freedom to accommodate residual data--simulation differences. The time-independent part for the continuum is described with empirical parametrisations determined from data. Statistical dependences among observables are included through conditional PDFs.

The time-dependent part is written as a PDF for $(t_{\rm tag}, q)$, conditional on $(\sigma_{t_{\rm tag}}, w_t)$, multiplied by a PDF for $w_t$ and a PDF for $\sigma_{t_{\rm tag}}$ conditional on $M_{\rm bc}$ . This factorisation accounts for the different $\sigma_{t_{\rm tag}}$ distributions of signal, $B\overline{B}$, and continuum events.

The signal PDF for $(t_{\rm tag}, q)$ is obtained from  Eq.~(\ref{eq:prob_ft}) by convolving it with the decay-time resolution function of Eq.~(\ref{eq:res}). In this PDF, only $S_{\pi^0\pi^0}$ and $C_{\pi^0\pi^0}$ are free parameters; external quantities such as the $B^0$ lifetime, the $B^0$--$\overline{B}{}^0$ mixing frequency, and flavour-tagging calibration constants are Gaussian-constrained to their known values. The PDF for $\sigma_{t_{\rm tag}}$ is determined from simulation.

The $B\overline{B}$ background uses the same time-dependent structure as the signal, with effective parameters describing its lifetime and asymmetry constrained from simulation and from the time-integrated analysis~\cite{BelleIIpi0pi0BR}. Since the sample is dominated by $B^+ \to \rho^+\pi^0$ decays, we assume no time-dependent asymmetry and a systematic uncertainty is assigned to account for a possible effect from the small neutral-$B\overline{B}$ component.

Because continuum dominates the selected sample, its time-dependent description is determined with maximal flexibility from data. The~shape parameters of its $t_{\rm tag}$ distribution are freely determined by the fit, the $\sigma_{t_{\rm tag}}$ distribution is modelled as a function of $M_{\rm bc}$ using data, and a possible instrumental flavour asymmetry is absorbed by an effective asymmetry parameter determined by the fit.

\subsection*{Analysis of $B^0 \to J/\psi K_S^0$ decays}

The decay $B^0 \to J/\psi K^0_{\rm S}$ provides a high-yield, low-background control channel, with time-dependent \textit{CP}-violating coefficients $S_{J/\psi K^{0}_{\rm S}}$ and $C_{J/\psi K^{0}_{\rm S}}$ known with high precision~\cite{HFLAV:2022esi}, and therefore offers a stringent validation of the $t_{\rm tag}$-based approach.

We reconstruct $B^0 \to J/\psi K^0_{\rm S}$ decays following Ref.~\cite{flavorTagger}. 
 We form $J/\psi$ candidates from opposite-sign electron or muon pairs with invariant masses consistent with the known $J/\psi$ mass, using looser requirements for electrons to account for bremsstrahlung. Electrons and muons are selected using particle-identification information from all subdetectors except the pixel detector. We reconstruct $K^0_{\rm S}$ candidates from pairs of oppositely charged pions with invariant mass consistent with the known $K^0_{\rm S}$ mass and a decay vertex significantly displaced from the $e^+e^-$ interaction point. We fit the trajectories and momenta of the $B^0$ decay products, constraining the $J/\psi$ to its known mass~\cite{ParticleDataGroup} and the $B^0$ to originate from the interaction region.  In addition, each $B^0$ candidate must satisfy $M_{\rm bc} > 5.27~{\rm GeV}/c^2$, \mbox{$-0.1 < \Delta E < 0.25~{\rm GeV}$} and $R_2 < 0.4$~\cite{FoxWolfram1978}. Relative to Ref.~\cite{flavorTagger}, the only additional requirements are $|t_{\rm tag}| < 15~\mathrm{ps}$ and $0.35 < \sigma_{t_{\rm tag}} < 4.00~\mathrm{ps}$.

\begin{figure}[tb]
 \centering
   \includegraphics[width=180pt]{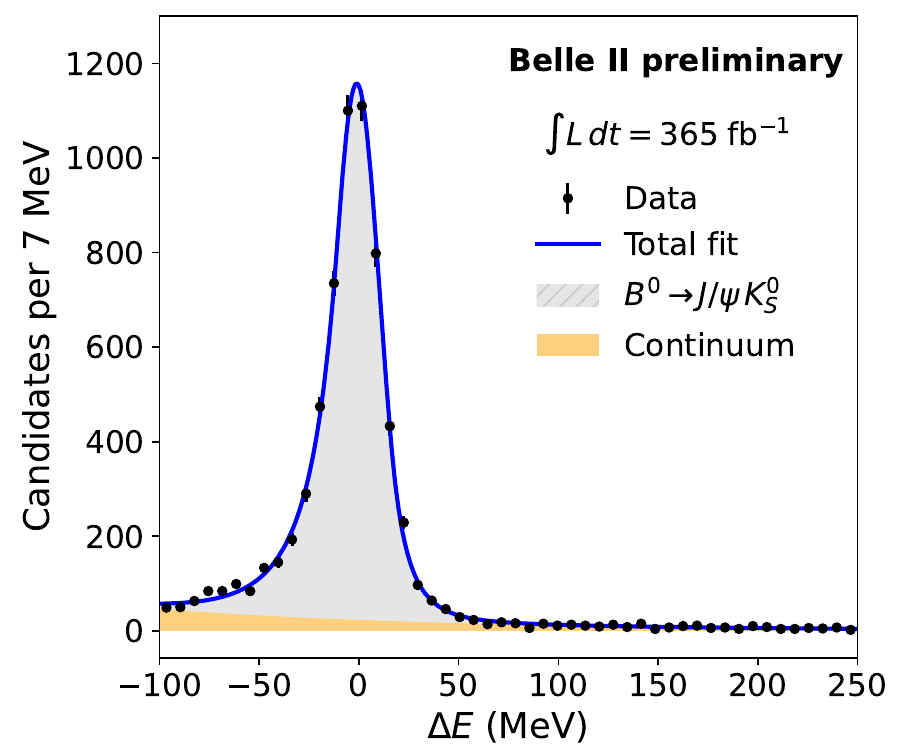}
 \caption{Distribution of $\Delta E$ for $B^0 \to J/\psi K_S^0$ candidates reconstructed in data, with fit projection overlaid. }
 \label{fig:jpsi_sweight}
\end{figure}

We first fit the unbinned $\Delta E$ distribution, shown in  Fig.~\ref{fig:jpsi_sweight}, to determine the sample composition. The resulting $s$Weights are then used in a second fit to the time evolution of background-subtracted, flavour-tagged signal candidates, shown in Fig.~\ref{fig:fitProjections-control} (right).

In this fit, we use the decay-time resolution function and flavour-tagging calibration obtained from the $B^0 \to D^{(*)-}\pi^+$ control samples. We measure \mbox{$S_{J/\psi K^{0}_{\rm S}}=0.835\pm0.104(\mathrm{stat})$} and \mbox{$C_{J/\psi K^{0}_{\rm S}}=-0.056\pm0.033(\mathrm{stat})$}, consistent with $S_{J/\psi K^{0}_{\rm S}}=0.724\pm0.035(\mathrm{stat})$ and \mbox{$C_{J/\psi K^{0}_{\rm S}}=-0.035\pm0.026(\mathrm{stat})$} from the standard $\Delta t$-based analysis on the same data~\cite{flavorTagger}, and with the known values~\cite{HFLAV:2022esi}.  
The linear correlation between the $S_{J/\psi K^{0}_{\rm S}}$ results from the two methods is 33\% (and 77\% for $C_{J/\psi K^{0}_{\rm S}}$). 
This measurement provides the first observation of mixing-induced \textit{CP} violation with a $t_{\rm tag}$-based analysis.

\subsection*{Systematic uncertainties}

Systematic uncertainties arise from modelling assumptions in the fit, potential biases in the extraction of the observables, and uncertainties in external inputs. Their contributions to $S_{\pi^0\pi^0}$ and $C_{\pi^0\pi^0}$ are summarised in  Tab.~\ref{table:uncertainty}.

\begin{table*}[tb]
\caption{Summary of systematic uncertainties on the $S_{\pi^0\pi^0}$ and $C_{\pi^0\pi^0}$ asymmetries. Total systematic uncertainties, resulting from their sums in quadrature, are also given.}
\label{table:uncertainty}
\centering
\begin{tabular}{l c c c c}
\toprule
 Source & $\phantom{<}\sigma(S_{\pi^0\pi^0})$ & \phantom{Source} & $\phantom{<}\sigma(C_{\pi^0\pi^0})$\\
\hline
$t_{\rm tag}$ resolution model &  $\phantom{<}0.08$  & & $<\!0.01$ \\
Fit bias & $\phantom{<}0.06$  & & $\phantom{<}0.04$ \\
$B\overline{B}$ background \textit{CP} violation  &  $\phantom{<}0.03$  & & $\phantom{<}0.03$ \\
$B\overline{B}$ background lifetime &  $<\!0.01$  & & $<\!0.01$\\
$\sigma_{t_{\rm tag}}$ modelling for continuum& $\phantom{<}0.01$ &  & $<\!0.01$ \\
$\sigma_{t_{\rm tag}}$ modelling for signal and $B\overline{B}$ decays & $\phantom{<}0.01$  & & $<\!0.01$  \\
Flavour-tagging parameters  &  $\phantom{<}0.01$ & & $\phantom{<}0.01$ \\
Flavour-tagging calibration model &  $\phantom{<}0.02$ &  & $\phantom{<}0.01$ \\
$\tau$ and $\Delta m$ uncertainties & $<\!0.01$ & & $<\!0.01$\\
Tag-side interference & $\phantom{<}0.01$ & & $\phantom{<}0.04$ \\
Continuum time-independent PDF modelling & $<\!0.01$ & & $<\!0.01$ \\
Signal time-independent PDF modelling  & $\phantom{<}0.03$ & & $\phantom{<}0.01$ \\
$B\overline{B}$ time-independent PDF modelling  & $\phantom{<}0.01$ & & $<\!0.01$ \\
\hline
Total systematic uncertainty &  $\phantom{<}0.11$  & & $\phantom{<}0.07$  \\
\bottomrule
\end{tabular} 
\end{table*}

Uncertainties associated with nuisance parameters constrained in the likelihood are included in the statistical uncertainties returned by the fit. The impact of each individual uncertainty is evaluated by repeating the fit with that parameter fixed to its best-fit value and computing the corresponding change in the uncertainties on $S_{\pi^0\pi^0}$ and $C_{\pi^0\pi^0}$.

The dominant systematic uncertainty arises from the modelling of the $t_{\rm tag}$ resolution function, in particular from possible differences between the $B^0 \to D^- \pi^+$ control sample and the signal sample as seen in simulation (see  Fig.~\ref{fig:resolutionFunction_Dpi_MC}).  A small difference arises because the tag-$B$ momentum is inferred using momentum conservation and therefore depends on the signal $B$ momentum, which is less precisely measured in the four-photon signal final state than in control modes with charged particles. We estimate the impact of this difference from alternative fits using models derived from the signal simulation.

Possible biases on $S_{\pi^0\pi^0}$ and $C_{\pi^0\pi^0}$ in the nominal fit are evaluated with ensembles of simulated experiments and are found to be small; we assign as a systematic uncertainty the combination in quadrature of the bias and its statistical precision.

Possible \textit{CP} violation in the $B\overline{B}$ background is assessed by introducing a time-dependent asymmetry for the small $B^0$ component and evaluating the resulting bias on the results; possible direct \textit{CP} violation is accounted for with a nuisance asymmetry parameter estimated in a $\Delta E$ sideband~\cite{BelleIIpi0pi0BR} and constrained in the fit. The uncertainty associated with the effective lifetime of the $B\overline{B}$ component is negligible. Uncertainties related to the modelling of $\sigma_{t_{\rm tag}}$ are evaluated separately for continuum, signal, and $B\overline{B}$ components using alternative PDFs derived from sidebands or control samples.

Flavour-tagging uncertainties are evaluated by varying the calibration parameters within their uncertainties and by testing an alternative calibration model with higher-order dependence on $w$. We also evaluate the effect of possible tag-side interference following Ref.~\cite{Long:2003wq}. Uncertainties associated with external physics parameters have negligible impact. Residual uncertainties related to continuum modelling and to correlations among observables are also small.

\subsection*{Isospin analysis}

We determine the \textit{CP}-violating phase $\phi_2$ from $B \to \pi\pi$ decays following Ref.~\cite{Gronau:1990ka}, using the isospin relations among amplitudes to separate the contributions of tree (single $W$ emission) and penguin ($W$ emission and reabsorption) topologies. The method relies on the approximate isospin symmetry of the strong interaction, which allows $\phi_2$ to be determined with negligible hadronic uncertainty.

In the isospin limit, the amplitudes $A^{ij}$ for $B \to \pi^i\pi^j$ decays, comprising $B^0 \to \pi^+\pi^-$, $B^0 \to \pi^0\pi^0$, and $B^\pm \to \pi^\pm\pi^0$, satisfy the triangle relations
\begin{align}
A^{+0} - A^{00} = \frac{1}{\sqrt{2}} A^{+-}\,, \\
\bar{A}^{+0} - \bar{A}^{00} = \frac{1}{\sqrt{2}} \bar{A}^{+-}\,,
\end{align}
where $\bar{A}^{ij}$ denote the amplitudes for $\overline{B}$ decays. The $B$ and $\overline{B}$ triangles share a common base given by the $B^\pm \to \pi^\pm\pi^0$ amplitude.

The measured observables constrain the amplitudes through the branching fractions $B_{ij}$ and, for the neutral modes, the \textit{CP}-violating coefficients $C_{ij}$ and $S_{ij}$:
\begin{align}
\frac{B_{ij}}{\tau_B} = \frac{|A^{ij}|^2 + |\bar A^{ij}|^2}{2}\,, \\
C_{ij} = \frac{|A^{ij}|^2 - |\bar A^{ij}|^2}{|A^{ij}|^2 + |\bar A^{ij}|^2}\,, \\
S_{ij} = \frac{2\,\mathrm{Im}(\bar A^{ij} A^{ij*})}{|A^{ij}|^2 + |\bar A^{ij}|^2}\,,
\end{align}
where $\tau_B$ indicates $\tau_{B^0}$ for $ij=+-,00$ and $\tau_{B^+}$ for $ij=+0$. We use the measurements listed in  Tab.~\ref{tab:alphaInputs}, including the $S_{\pi^0\pi^0}$ and $C_{\pi^0\pi^0}$ results reported here.

\begin{table*}[tb]
\caption{Experimental inputs used in the determination of the allowed $\phi_2$  intervals. The leftmost column shows the relevant quantities; the second column from left shows each result as reported in the original publications; the third column reports the experiment associated with those results; the fourth column reports the average of measurements as reported by the  HFLAV group~\cite{HFLAV}; the rightmost column reports the averages scaled to account for updated values of $f^{+-}$ and $f^{00}$~\cite{HFLAV:2024ctg} for CLEO, Belle, and Babar measurements, and of $\mathcal{B}(B^0\to K^+\pi^-)$ for LHCb and CDF~\cite{ParticleDataGroup}, which use it to normalise $\mathcal{B}(B^0\to \pi^+\pi^-)$. The third contribution to the uncertainty in the scaled averages arises from the uncertainties on $f^{+-}$ and $f^{00}$. These are symmetrised by taking the larger asymmetric uncertainty and are treated as uncorrelated, since the correlation is negligible after symmetrisation. The $\mathcal{C}_{\pi^0\pi^0}$ value reported for Belle II in square brackets is the time-integrated measurement, which is superseded by the results of this work and therefore removed from the inputs of the intervals that include our results; the average in square brackets contains that value and is used only as input to the intervals that do not include our measurements.}
\label{tab:alphaInputs}
\centering
\begin{tabular}{lllcc}
\toprule
Observable &  $\phantom{-}$Input measurements & Experiment & HFLAV  & Rescaled average  \\
           &                     &   & average~\cite{HFLAV}      & (our computation) \\
\hline
\multirow{4}{*}{$ B_{\pi^+\pi^0}$ $[10^{-6}]$}   &$\phantom{-}5.86 \pm 0.26 \pm 0.38$ &Belle~\cite{Br+0Belle} & \multirow{4}{*}{$5.33\pm 0.27$}  & \multirow{4}{*}{$5.268 \pm 0.258 \pm 0.111$} \\
                              &$\phantom{-}5.10 \pm 0.29 \pm 0.27$ &Belle~II~\cite{Br+0BelleII} &       &  \\
                               &$\phantom{-}5.02 \pm0.46 \pm 0.29$ &Babar~\cite{Br+0Babar} &       &  \\
                               &$\phantom{-}{4.6}^{+1.8}_{-1.6}{}^{+0.6}_{-0.7}$ &CLEO~\cite{Br+0Cleo} &       &  \\
                              \hline
\multirow{6}{*}{$B_{\pi^+\pi^-}$ $[10^{-6}]$}  &$\phantom{-}5.26 \pm 0.18 \pm 0.36$ &LHCb~\cite{Br+-LHCb} &\multirow{6}{*}{$5.36\pm 0.16$} & \multirow{6}{*}{$5.467 \pm 0.150 \pm 0.067$} \\
                               &$\phantom{-}5.83 \pm 0.22 \pm 0.17$ &Belle~II~\cite{Br+0BelleII} &       &  \\
                               &$\phantom{-}5.04 \pm0.21 \pm0.18$ &Belle~\cite{Br+0Belle} &       &  \\
                               &$\phantom{-}5.20 \pm0.34 \pm0.34$ &CDF~\cite{Br+-CDF} &       &  \\
                               &$\phantom{-}5.5 \pm0.4 \pm0.3$ &Babar~\cite{Br+-BaBar} &       &  \\
                               &$\phantom{-}4.5 ^{+1.4} _{-1.2}{}^{+0.5} _{-0.4}$ &CLEO~\cite{Br+0Cleo} &       &  \\
                              \hline
\multirow{3}{*}{$B_{\pi^0\pi^0}$ $[10^{-6}]$}   &$\phantom{-}1.31 \pm0.19 \pm0.19$ &Belle~\cite{Br00Belle} & \multirow{3}{*}{$1.46\pm 0.14$} & \multirow{3}{*}{$1.482 \pm 0.145 \pm 0.024$} \\
                              &$\phantom{-}1.83 \pm0.21 \pm0.13$ &Babar~\cite{Br00BaBar} &       &  \\
                               &$\phantom{-}1.25 \pm0.20 \pm0.11$ &Belle~II~\cite{BelleIIpi0pi0BR} &       &  \\
                              \hline
 \multirow{3}{*}{$C_{\pi^+\pi^-}$}  &$-0.25 \pm 0.08 \pm 0.02$ &Babar~\cite{Br00BaBar} &\multirow{3}{*}{ $-0.311\pm 0.030$ } &\\
                 &$-0.33 \pm 0.06 \pm 0.03$ &Belle~\cite{BelleSC+-} &  & \\
                 &$-0.320 \pm 0.038$ &LHCb~\cite{LHCbCS+-} & & \\
                \hline
\multirow{3}{*}{$S_{\pi^+\pi^-}$}  &$-0.68 \pm 0.10 \pm 0.03$ &Babar~\cite{Br00BaBar} &\multirow{3}{*}{$-0.666\pm 0.029 $} &\\
                 &$-0.64 \pm 0.08 \pm 0.03$ &Belle~\cite{BelleSC+-} &            & \\
                 &$-0.672 \pm 0.034$ &LHCb~\cite{LHCbCS+-} & & \\
                \hline
\multirow{3}{*}{$C_{\pi^0\pi^0}$}  &$-0.43 \pm0.26 \pm0.05$ &Babar~\cite{Br00BaBar} &\multirow{2}{*}{$-0.33\pm 0.22$} & \\
                   &$-0.14 \pm 0.36 \pm 0.10$ &Belle~\cite{Br00Belle} &  &  \\
                   &[$-0.03 \pm0.30 \pm0.04$] &Belle~II~\cite{BelleIIpi0pi0BR} & [$-0.23 \pm 0.18$]  & \\
\hline
$C_{\pi^0\pi^0}$  &$\phantom{-}0.05 \pm 0.28 \pm 0.07$ & This work & & \\
$S_{\pi^0\pi^0}$  &$\phantom{-}0.61_{-0.79}^{+0.75} \pm 0.11$  & This work & & \\
\toprule
\end{tabular}
\end{table*}

Following Ref.~\cite{Charles:2017evz}, we parameterise the amplitudes as
\begin{align}
A^{+0}\, &= \mu\, e^{i(\Delta-\phi_2)}, &
\bar A^{+0}\, &= \mu\, e^{i(\Delta+\phi_2)}, \nonumber\\
A^{+-} &= \mu\, a, &
\bar A^{+-} &= \mu\, \bar a\, e^{2i\bar\phi_2}, \nonumber\\
A^{00}\,\, &= A^{+0} - \frac{A^{+-}}{\sqrt{2}}, &
\bar A^{00}\,\, &= \bar A^{+0} - \frac{\bar A^{+-}}{\sqrt{2}},
\end{align}
where $\Delta$ is a strong phase, and $a$, $\bar a$, $\mu$, and $\bar\phi_2$ are real parameters. This parametrisation relates the seven observables of the $B \to \pi\pi$ system to six physics parameters.

\begin{figure}[tb]
\centering
\includegraphics[width=220pt]{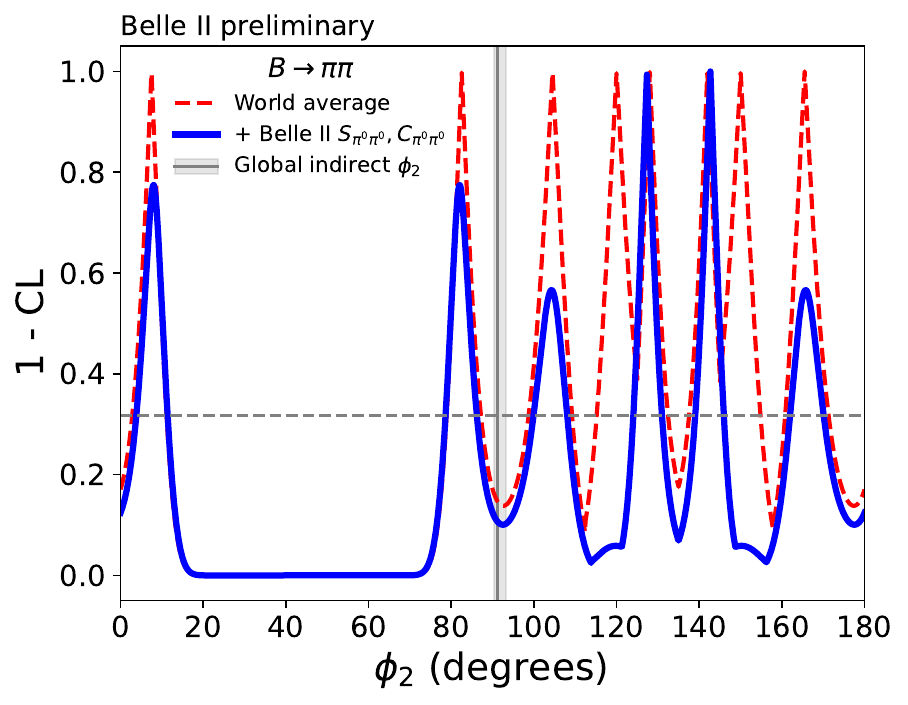}
\caption{Values of $(1-\mathrm{CL)}$ as a function of $\phi_2$ as determined using $B\to \pi\pi$ decays (dashed red line) before and (solid blue line) after including our measurements of $S_{\pi^0\pi^0}$ and $C_{\pi^0\pi^0}$. The vertical line shows the indirect value of $\phi_2$  obtained from the global CKM fit. The horizontal dashed line represents 68.3\%  confidence-level exclusion.} 
\label{fig:alphaScan}
\end{figure}

For each value of $\phi_2$, we construct
\begin{equation}
\chi^2 = (\vec{O} - \vec{O}_p)^T V^{-1} (\vec{O} - \vec{O}_p),
\end{equation}
where $\vec{O}$ is the vector of measured observables, $\vec{O}_p$ the corresponding prediction, and $V$ the covariance matrix including experimental uncertainties and correlations. The $\chi^2$ is numerically minimised with respect to the remaining parameters $(\bar\phi_2, a, \bar a, \Delta, \mu)$, and the resulting profiled $\chi^2_p(\phi_2)$ is used to construct confidence intervals from
\begin{equation}
\Delta \chi^2 = \chi^2_p(\phi_2) - \chi^2_0(\hat\phi_2),
\end{equation}
where $\hat\phi_2$ is the value of $\phi_2$ at the global minimum. The allowed $\phi_2$ interval at the 68.3\% CL is
\begin{equation}
\begin{split}
    [3.8^\circ, 11.4^\circ]    & \cup  [78.6^\circ, 86.2^\circ]    \cup  [99.9^\circ, 108.0^\circ]\, \cup \\ 
    [124.1^\circ, 131.0^\circ] & \cup  [139.0^\circ, 145.9^\circ]  \cup  [162.0^\circ, 170.1^\circ]
\end{split}
\end{equation}
 and is shown in  Fig.~\ref{fig:alphaScan}.
 
 We use an analogous procedure to obtain the two-dimensional confidence regions in the $(S_{\pi^0\pi^0},\phi_2)$ plane shown in Fig.~\ref{fig:alpha_vs_S00}. We express one nuisance parameter ($a$) as a function of $S_{\pi^0\pi^0}$ and the other parameters, and, for each fixed $(S_{\pi^0\pi^0},\phi_2)$ pair, we numerically minimise the $\chi^2$ with respect to the remaining parameters, obtaining the profiled value $\chi^2_p(S_{\pi^0\pi^0},\phi_2)$. The $68.3\%$ CL regions are defined by the contour $\Delta\chi^2=2.3$, where $\Delta\chi^2$ is the difference between the profiled $\chi^2$ and its value at the global minimum.

Isospin-breaking effects, such as electroweak penguin contributions and $\pi^0$--$\eta^{(\prime)}$ mixing, are expected to induce corrections at the level of $1^\circ$--$2^\circ$~\cite{Charles:2017evz} and are neglected in this analysis.

\backmatter

\bmhead{Acknowledgements}
% Policy from October 20, 2022
This work, based on data collected using the Belle II detector, which was built and commissioned prior to March 2019,
%Belle1 and data collected using the Belle detector, which was operated until June 2010,
was supported by
%Armenia
Higher Education and Science Committee of the Republic of Armenia Grant No.~23LCG-1C011;
%Australia
Australian Research Council and Research Grants
No.~DP200101792, % Jackson
No.~DP210101900, % Urquijo
No.~DP210102831, % Sevior
No.~DE220100462, % Hsu
No.~LE210100098, % Infrastructure
and
No.~LE230100085; % Infrastructure
%Austria
Austrian Federal Ministry of Education, Science and Research,
Austrian Science Fund (FWF) Grants
DOI:~10.55776/P34529,
DOI:~10.55776/J4731,
DOI:~10.55776/J4625,
DOI:~10.55776/M3153,
and
DOI:~10.55776/PAT1836324,
and
Horizon 2020 ERC Starting Grant No.~947006 ``InterLeptons'';
%Canada
Natural Sciences and Engineering Research Council of Canada, Digital Research Alliance of Canada, and Canada Foundation for Innovation;
%China
National Key R\&D Program of China under Contract No.~2024YFA1610503,
and
No.~2024YFA1610504
National Natural Science Foundation of China and Research Grants
No.~11575017,
No.~11761141009,
No.~11705209,
No.~11975076,
No.~12135005,
No.~12150004,
No.~12161141008,
No.~12405099,
No.~12475093,
and
No.~12175041,
and Shandong Provincial Natural Science Foundation Project~ZR2022JQ02;
%Czech Republic
the Czech Science Foundation Grant No. 22-18469S,  Regional funds of EU/MEYS: OPJAK
FORTE CZ.02.01.01/00/22\_008/0004632 
and
Charles University Grant Agency project No. 246122;
%EU
European Research Council, Seventh Framework PIEF-GA-2013-622527,
Horizon 2020 ERC-Advanced Grants No.~267104 and No.~884719,
Horizon 2020 ERC-Consolidator Grant No.~819127,
Horizon 2020 Marie Sklodowska-Curie Grant Agreement No.~700525 ``NIOBE''
and
No.~101026516,
and
Horizon Europe Marie Sklodowska-Curie Staff Exchange project JENNIFER3 Grant Agreement No.~101183137 (European grants);
%France
L’Institut National de Physique Nucl\'eaire et de Physique des
Particules (IN2P3) du CNRS under Project Identification No.
CNRS-IN2P3-14-PP-033
and L’Agence Nationale de la Recherche (ANR) under Grant No. ANR-23-CE31-
0018 and ANR-25-CE31-1333 (France);
%Germany
BMFTR, DFG, HGF, MPG, and AvH Foundation (Germany);
%India
Department of Atomic Energy under Project Identification No.~RTI 4002,
Department of Science and Technology,
and
UPES SEED funding programs
No.~UPES/R\&D-SEED-INFRA/17052023/01 and
No.~UPES/R\&D-SOE/20062022/06 (India);
%Israel
Israel Science Foundation Grant No.~2476/17,
U.S.-Israel Binational Science Foundation Grant No.~2016113, and
Israel Ministry of Science Grant No.~3-16543;
%Italy
Istituto Nazionale di Fisica Nucleare and the Research Grants BELLE2,
and
the ICSC – Centro Nazionale di Ricerca in High Performance Computing, Big Data and Quantum Computing, funded by European Union – NextGenerationEU;
%Japan
Japan Society for the Promotion of Science, Grant-in-Aid for Scientific Research Grants
No.~16H03993,
No.~16H06492,
No.~16K05323,
No.~17H01133,
No.~17H05405,
No.~18K03621,
No.~18H03710,
No.~18H05226,
No.~19H00682, % Niigata
No.~20H05850,
No.~20H05858,
No.~22H00144,
No.~22K14056,
No.~22K21347,
No.~23H05433,
No.~26220706,
No.~26400255,
and
No.~26H02056,
%the National Institute of Informatics, and Science Information NETwork 5 (SINET5), 
and
the Ministry of Education, Culture, Sports, Science, and Technology (MEXT) of Japan;  
%Korea
National Research Foundation (NRF) of Korea Grants
No.~2021R1-F1A-1064008,
No.~2022R1-A2C-1003993,
No.~RS-2018-NR031074,
No.~RS-2021-NR060129,
No.~RS-2024-00354342,
No.~RS-2025-02219521,
No.~RS-2026-25471491,
No.~RS-2026-25480677,
and
No.~RS-2026-25486791,
Radiation Science Research Institute,
Foreign Large-Size Research Facility Application Supporting project,
the Global Science Experimental Data Hub Center, the Korea Institute of Science and
Technology Information (K26L1M2C3)
and
KREONET/GLORIAD;
%Malaysia
Universiti Malaya RU grant, Akademi Sains Malaysia, and Ministry of Education Malaysia;
%Mexico
% CINVESTAV-IPN, UNAM, UAS, BUAP and CONACYT are funded under
Frontiers of Science Program Contracts
No.~FOINS-296,
No.~CB-221329,
No.~CB-236394,
No.~CB-254409,
and
No.~CB-180023, and SEP-CINVESTAV Research Grant No.~237 (Mexico);
%Poland
the Polish Ministry of Science and Higher Education and the National Science Center;
%Russia
the Ministry of Science and Higher Education of the Russian Federation
and
the HSE University Basic Research Program, Moscow;
%Saudi Arabia
University of Tabuk Research Grants
No.~S-0256-1438 and No.~S-0280-1439 (Saudi Arabia);
%Slovenia
Slovenian Research Agency and Research Grants
No.~J1-50010
and
No.~P1-0135;
%Spain
Ikerbasque, Basque Foundation for Science,
State Agency for Research of the Spanish Ministry of Science and Innovation through Grant No. PID2022-136510NB-C33, Spain,
Agencia Estatal de Investigacion, Spain
Grant No.~RYC2020-029875-I
and
Generalitat Valenciana, Spain
Grant No.~CIDEGENT/2018/020;
%Swiss (Belle 1)
%Belle1 the Swiss National Science Foundation;
%Sweden
The Knut and Alice Wallenberg Foundation (Sweden), Contracts No.~2021.0174, No.~2021.0299, and No.~2023.0315;
%Taiwan
National Science and Technology Council,
and
Ministry of Education (Taiwan);
%Thailand
Thailand Center of Excellence in Physics;
%Turkey
TUBITAK ULAKBIM (Turkey);
%Ukraine
National Research Foundation of Ukraine, Project No.~2020.02/0257,
and
Ministry of Education and Science of Ukraine;
%USA
the U.S. National Science Foundation and Research Grants
No.~PHY-1913789 % Indiana CEEM
and
No.~PHY-2111604, % Luther
and the U.S. Department of Energy and Research Awards
No.~DE-AC06-76RLO1830, % PNNL
No.~DE-SC0007983, % Wayne State
No.~DE-SC0009824, % Florida
No.~DE-SC0009973, % VPI
No.~DE-SC0010007, % Duke
No.~DE-SC0010073, % South Carolina
No.~DE-SC0010118, % Carnegie Mellon
No.~DE-SC0010504, % Hawaii
No.~DE-SC0011784, % Cincinnati
No.~DE-SC0012704, % BNL
No.~DE-SC0019230, % Duke
No.~DE-SC0021616, % Mississippi
No.~DE-SC0022350, % Louisville
No.~DE-SC0023470; % South Alabama
%last group
and
%Vietnam
the Vietnam Academy of Science and Technology (VAST) under Grant
No.~DL0000.05/26-27.

% Policy from October 20, 2022
These acknowledgements are not to be interpreted as an endorsement of any statement made
by any of our institutes, funding agencies, governments, or their representatives.

We thank the SuperKEKB team for delivering high-luminosity collisions;
the KEK cryogenics group for the efficient operation of the detector solenoid magnet and IBBelle on site;
the KEK Computer Research Center for on-site computing support; the NII for SINET6 network support;
and the raw-data centers hosted by BNL, DESY, GridKa, IN2P3, INFN, 
%Belle1 PNNL/EMSL, 
and the University of Victoria.

%Please refer to Journal-level guidance for any specific requirements.

%\section*{Declarations}

%Some journals require declarations to be submitted in a standardised format. Please check the Instructions for Authors of the journal to which you are submitting to see if you need to complete this section. If yes, your manuscript must contain the following sections under the heading `Declarations':

%\begin{itemize}
%\item Funding
%\item Conflict of interest/Competing interests (check journal-specific guidelines for which heading to use)
%\item Ethics approval and consent to participate
%\item Consent for publication
%\item Data availability 
%\item Materials availability
%\item Code availability 
%\item Author contribution
%\end{itemize}

%\noindent
%If any of the sections are not relevant to your manuscript, please include the heading and write `Not applicable' for that section. 

%%===========================================================================================%%
%% If you are submitting to one of the Nature Portfolio journals, using the eJP submission   %%
%% system, please include the references within the manuscript file itself. You may do this  %%
%% by copying the reference list from your .bbl file, paste it into the main manuscript .tex %%
%% file, and delete the associated \verb+\bibliography+ commands.                            %%
%%===========================================================================================%%

\bibliography{sn-bibliography}% common bib file

@article{BelleIIpi0pi0BR,
  title = {Measurement of the branching fraction and {$CP$}-violating asymmetry of the decay ${B}^{0}\ensuremath{\rightarrow}{\ensuremath{\pi}}^{0}{\ensuremath{\pi}}^{0}$ using 387 million $\mathrm{\ensuremath{\Upsilon}}{(4S)}$ decays in {Belle~II} data},
  author = {Adachi, I. and others},
  journal = {Phys. Rev. D},
  volume = {111},
  issue = {7},
  pages = {L071102},
  year = {2025}
}

@misc{abe2010belleiitechnicaldesign,
      title={Belle II Technical Design Report}, 
      author={T. Abe and others},
      year={2010},
      note="Preprint at \url{https://arxiv.org/abs/1011.0352}" 
}

@article{FlavorTagger,
  title = {New graph-neural-network flavor tagger for {Belle~II} and measurement of {sin $2{\ensuremath{\phi}}_{1}$} in {${B}^{0}\ensuremath{\rightarrow}J/\ensuremath{\psi}{K}_{\mathrm{S}}^{0}$} decays},
  author = {Adachi, I. and others},
  journal = {Phys. Rev. D},
  volume = {110},
  issue = {1},
  pages = {012001},
  year = {2024}
}

@article{Charles:2017evz,
    author = "Charles, J. and Deschamps, O. and Descotes-Genon, S. and Niess, V.",
    title = "{Isospin analysis of charmless {\it{B}}-meson decays}",
    journal = "Eur. Phys. J. C",
    volume = "77",
    pages = "574",
    year = "2017"
}

@article{Gronau:1990ka,
    author = "Gronau, Michael and London, David",
    title = "{Isospin analysis of {\it{CP}} asymmetries in {\it{B}} decays}",
    journal = "Phys. Rev. Lett.",
    volume = "65",
    pages = "3381",
    year = "1990"
}

@article{PhysRevD.60.111301,
  title = {{Measurement of \textit{CP} violation at the {$\ensuremath{\Upsilon}(4S)$} without time ordering or $\ensuremath{\Delta}t$}},
  author = {Foland, Andrew D.},
  journal = {Phys. Rev. D},
  volume = {60},
  issue = {11},
  pages = {111301},
  numpages = {5},
  year = {1999},
}

@article{rzr8-l6l8,
  title = {{Unlocking time-dependent \textit{CP} violation without signal vertexing at {$B$} factories}},
  author = {Dorigo, M. and Raiz, S. and Tonelli, D. and Žlebčík, R.},
  journal = {Phys. Rev. D},
  volume = {112},
  issue = {3},
  pages = {032011},
  numpages = {10},
  year = {2025},
}

@article{AKAI2018188,
title = {{SuperKEKB collider}},
journal = {Nucl. Instrum. Methods Phys. Res. A},
volume = {907},
pages = {188},
year = {2018},
author = {Kazunori Akai and Kazuro Furukawa and Haruyo Koiso},
}

@article{Miyabayashi_2020,
year = {2020},
volume = {15},
pages = {C10016},
author = {Miyabayashi, K.},
title = {{Belle II electromagnetic calorimeter and its performance during early SuperKEKB operation}},
journal = {J. Instrum.},
}

@book{Fisher,
  author		= "R. A. Fisher",
  title			= "{Statistical Methods and Scientific Inference}",
  address		= "Edinburgh",
  publisher		= "{Oliver and Boyd}",
  year			= "1959"
}

@article{Belle-II:2018jsg,
    author = "Altmannshofer, W. and others",
    title = "{{The Belle II Physics Book}}",
    journal = "PTEP",
    volume = "2019",
    pages = "123C01",
    year = "2019",
    note = "[Erratum: PTEP 2020, 029201 (2020)]"
}

@article{10.1143/PTP.49.652,
    author = {Kobayashi, Makoto and Maskawa, Toshihide},
    title = {{\textit{CP}-Violation in the Renormalizable Theory of Weak Interaction}},
    journal = {Progr. Theor. Phys.},
    volume = {49},
    pages = {652},
    year = {1973},
}

@article{PIVK2005356,
title = {{$s$Plot: a statistical tool to unfold data distributions}},
author = {M. Pivk and F.R. {Le Diberder}},
journal = {{Nucl. Instrum. Methods Phys. Res. A}},
volume = {555},
pages = {356},
year = {2005},
}

@article{HFLAV,
    author = "Banerjee, Swagato and others",
    journal = {Phys. Rev.},
    volume = {D113},
    issue = {1},
    pages = {012008},
    title = "{{Averages of $b$-hadron, $c$-hadron, and $\tau$-lepton properties as of 2023}}",
    year = "2026"
}

@article{HFLAV:2024ctg,
    author = "Banerjee, Swagato and others",
    journal = {Phys. Rev.},
    volume = {D113},
    issue = {1},
    pages = {012008},
    title = "{{Averages of $b$-hadron, $c$-hadron, and $\tau$-lepton properties as of 2023}}",
    year = "2026",
    note = "{With specific results from \href{https://doi.org/10.5281/zenodo.19174281}{{\texttt{doi:10.5281/zenodo.19210609}}}}"
}

@article{ParticleDataGroup,
    author = "Navas, S. and others",
    title = "{Review of particle physics}",
    journal = "Phys. Rev. D",
    volume = "110",
    pages = "030001",
    year = "2024"
}

@article{Efron1979,
  author  = {Efron, Bradley},
  title   = {Bootstrap Methods: Another Look at the Jackknife},
  journal = {Ann. Statist.},
  volume  = {7},
  pages   = {1},
  year    = {1979}
}

@article{Belle-II:2024frs,
    author = "Adachi, I. and others",
    title = "{Measurement of the branching fraction, polarization, and time-dependent CP asymmetry in $B^0\rightarrow\rho^+\rho^-$ decays and constraint on the CKM angle $\phi_2$}",
    journal = "Phys. Rev. D",
    volume = "111",
    pages = "092001",
    year = "2025"
}

@article{PhysRevD.1.1416,
	title        = {Statistical Model for Electron-Positron Annihilation into Hadrons},
	author       = {Bjorken, James D. and Brodsky, Stanley J.},
	year         = 1970,
	journal      = {Phys. Rev. D},
	publisher    = {American Physical Society},
	volume       = 1,
	pages        = {1416},
	issue        = 5,
	numpages     = {0}
}

@article{Brandt:1964sa,
	title        = "{The Principal axis of jets. An Attempt to analyze high-energy collisions as two-body processes}",
	author       = "Brandt, S. and Peyrou, C. and Sosnowski, R. and Wroblewski, A.",
	year         = 1964,
	journal      = "Phys. Lett.",
	volume       = 12,
	pages        = "57"
}

@article{Farhi:1977sg,
	title        = "{A QCD Test for Jets}",
	author       = "Farhi, Edward",
	year         = 1977,
	journal      = "Phys. Rev. Lett.",
	volume       = 39,
	pages        = "1587"
}

@article{Long:2003wq,
    author = "Long, Owen and Baak, Max and Cahn, Robert N. and Kirkby, David P.",
    title = "{Impact of tag side interference on time dependent $C\!P$ asymmetry measurements using coherent $B^0 \overline{B}{}{^0}$ pairs}",
    journal = "Phys. Rev. D",
    volume = "68",
    pages = "034010",
    year = "2003"
}

@article{Keck:2017gsv,
	title        = "{FastBDT: A Speed-Optimized Multivariate Classification Algorithm for the Belle II Experiment}",
	author       = "Keck, Thomas",
	year         = 2017,
	journal      = "Comput. Softw. Big Sci.",
	volume       = 1,
	pages        = 2
}

@article{FoxWolfram1978,
  author    = {G. C. Fox and S. Wolfram},
  title     = "{Observables for the Analysis of Event Shapes in $e^+ e^-$ Annihilation and Other Processes}",
  journal   = {Phys. Rev. Lett.},
  volume    = {41},
  pages     = {1581},
  year      = {1978},
}

@article{Abe2001PRL,
  author    = {Abe, K. and others},
  title     = "{Measurement of Branching Fractions for $B \to \pi\pi, K\pi$ and $KK$ Decays}",
  journal   = {Phys. Rev. Lett.},
  volume    = {87},
  pages     = {101801},
  year      = {2001},
}

@article{Abe2001PLB,
  author    = {Abe, K. and others},
  title     = "{A measurement of the branching fraction for the inclusive $B \to X_s \gamma$ decays with the Belle detector}",
  journal   = {Phys. Lett. B},
  volume    = {511},
  pages     = {151},
  year      = {2001}
}

@article{Charles:2004jd,
    author = "Charles, J. and Hocker, Andreas and Lacker, H. and Laplace, S. and Le Diberder, F. R. and Malcles, J. and Ocariz, J. and Pivk, M. and Roos, L.",
    collaboration = "CKMfitter Group",
    title = "{CP violation and the CKM matrix: Assessing the impact of the asymmetric $B$ factories}",
    journal = "Eur. Phys. J. C",
    volume = "41",
    pages = "1",
    year = "2005",
    note = "{Updated results and plots available at: \href{http://ckmfitter.in2p3.fr}{{\texttt{http://ckmfitter.in2p3.fr}}}}"
}

@article{HFLAV:2022esi,
  author = "Amhis, Yasmine Sara and others",
  collaboration = "HFLAV",
  title = "{Averages of b-hadron, c-hadron, and {\ensuremath{\tau}}-lepton properties as of 2021}",
  journal = "Phys. Rev.",
  volume = "D107",
  pages = "052008",
  year = "2023",
  note = "{With specific results from \href{https://doi.org/10.5281/zenodo.19371140}{{\texttt{doi:10.5281/zenodo.19371140}}}}"
}

@article{Belle-IIDEPFET:2021pib,
  author = {Ye, H. and others},
  collaboration = {Belle-II DEPFET and PXD Collaborations},
  title = {{Commissioning and performance of the Belle~II pixel detector}},
  journal = {Nucl. Instrum. Meth. A},
  volume = {987},
  pages = {164875},
  year = {2021}
}

@article{Br+0Belle,
    author = "Duh, Y. -T. and others",
    collaboration = "Belle",
    title = "{Measurements of branching fractions and direct CP asymmetries for $B\to K\pi$, $B\to \pi\pi$ and $B\to KK$ decays}",
    journal = "Phys. Rev. D",
    volume = "87",
    pages = "031103",
    year = "2013"
}

@article{Br+0BelleII,
    author = "Adachi, I. and others",
    collaboration = "Belle-II",
    title = "{Measurement of branching fractions and direct CP asymmetries for $B\to K\pi$ and $B\to\pi\pi$ decays at Belle II}",
    journal = "Phys. Rev. D",
    volume = "109",
    pages = "012001",
    year = "2024"
}

@article{Br+0Babar,
    author = "Aubert, Bernard and others",
    editor = "Sissakian, Alexey and Kozlov, Gennady and Kolganova, Elena",
    collaboration = "BaBar",
    title = "{Observation of $B^{+} \to \bar{K}^0 K^{+}$ and $B^0 \to K^0 \bar{K}^0$}",
    journal = "Phys. Rev. Lett.",
    volume = "97",
    pages = "171805",
    year = "2006"
}

@article{Br+0Cleo,
    author = "Bornheim, A. and others",
    collaboration = "CLEO",
    title = "{Measurements of charmless hadronic two body $B$ meson decays and the ratio $\mathcal{B}(B \to D K) / \mathcal{B}(B \to D \pi)$}",
    journal = "Phys. Rev. D",
    volume = "68",
    pages = "052002",
    year = "2003",
    note = "[Erratum: Phys.Rev.D 75, 119907 (2007)]"
}

@article{Br+-LHCb,
    author = "Aaij, R and others",
    collaboration = "LHCb",
    title = "{Measurement of $b$-hadron branching fractions for two-body decays into charmless charged hadrons}",
    journal = "JHEP",
    volume = "10",
    pages = "037",
    year = "2012"
}

@article{Br+-CDF,
    author = "Aaltonen, T. and others",
    collaboration = "CDF",
    title = "{Measurements of Direct {\it CP} Violating Asymmetries in Charmless Decays of Strange Bottom Mesons and Bottom Baryons}",
    journal = "Phys. Rev. Lett.",
    volume = "106",
    pages = "181802",
    year = "2011"
}

@article{Br+-BaBar,
    author = "Aubert, Bernard and others",
    collaboration = "BaBar",
    title = "{Improved Measurements of the Branching Fractions for $B^0 \to \pi^{+} \pi^{-}$ and $B^0 \to K^{+} \pi^{-}$, and a Search for $B^0 \to K^{+} K^{-}$}",
    journal = "Phys. Rev. D",
    volume = "75",
    pages = "012008",
    year = "2007"
}

@article{Br00Belle,
    author = "Julius, T. and others",
    collaboration = "Belle",
    title = "{Measurement of the branching fraction and {\it CP} asymmetry in $B^{0} \to \pi^{0}\pi^{0}$ decays, and an improved constraint on $\phi_{2}$}",
    journal = "Phys. Rev. D",
    volume = "96",
    pages = "032007",
    year = "2017"
}

@article{Br00BaBar,
    author = "Lees, J. P. and others",
    collaboration = "BaBar",
    title = "{Measurement of {\it CP} Asymmetries and Branching Fractions in Charmless Two-Body $B$-Meson Decays to Pions and Kaons}",
    journal = "Phys. Rev. D",
    volume = "87",
    pages = "052009",
    year = "2013"
}

@article{LHCbCS+-,
    author = "Aaij, Roel and others",
    collaboration = "LHCb",
    title = "{Observation of {\it CP} violation in two-body $ {B}_{(s)}^0 $-meson decays to charged pions and kaons}",
    journal = "JHEP",
    volume = "03",
    pages = "075",
    year = "2021"
}

@article{BelleSC+-,
    author = "Adachi, I. and others",
    collaboration = "Belle",
    title = "{Measurement of the {\it CP} violation parameters in $B^0 \to \pi^+ \pi^-$ decays}",
    journal = "Phys. Rev. D",
    volume = "88",
    pages = "092003",
    year = "2013"
}

@article{Belle:2010igk,
    author = "Das, A. and others",
    collaboration = "Belle",
    title = "{Measurements of Branching Fractions for $B^0 \to D_s^+\pi^-$ and $\bar{B}^0 \to D_s^+K^-$}",
    journal = "Phys. Rev. D",
    volume = "82",
    pages = "051103",
    year = "2010"
}

@article{BaBar:2008xvo,
    author = "Aubert, Bernard and others",
    collaboration = "BaBar",
    title = "{Measurement of the Branching Fractions of the Rare Decays $B^0 \to D_s^{(*)}\pi^{-}$, $B^0 \to D_s^{(*)}\rho^{-}$, and $B^0 \to D_s^{(*)}K^{(*)+}$}",
    journal = "Phys. Rev. D",
    volume = "78",
    pages = "032005",
    year = "2008"
}

@article{LHCb:2020zae,
    author = "Aaij, R. and others",
    collaboration = "LHCb",
    title = "{Measurement of the branching fraction of the ${{B} ^0} {\rightarrow }{{D} ^+_{s}} {{\pi } ^-} $ decay}",
    journal = "Eur. Phys. J. C",
    volume = "81",
    pages = "314",
    year = "2021"
}
%% if required, the content of .bbl file can be included here once bbl is generated
%%\input sn-article.bbl

\end{document}